%% file: Paper_Revised_V3.tex
\documentclass[twocolumn,english,american,prb]{revtex4-1}
\usepackage[T1]{fontenc}
\usepackage[latin9]{inputenc}
\usepackage{amsmath}
\usepackage{amssymb}
\usepackage{graphicx}
\usepackage{babel}
\usepackage{xcolor}
\usepackage{natbib}

\usepackage{ulem}
\begin{document}
\title{Multigaps superconductivity at unconventional Lifshitz transition
in a 3D Rashba heterostructure at atomic limit}
\author{Maria Vittoria Mazziotti}
\affiliation{Dipartimento di Matematica e Fisica, Universit\`a  Roma Tre, via della
Vasca Navale 84 00146 Roma, Italy}
\author{Antonio Valletta}
\affiliation{Italian National Research Council CNR, Institute for Microelectronics and Microsystems IMM,
via del Fosso del Cavaliere, 100, 00133 Roma, Italy}

\author{Roberto Raimondi}
\affiliation{Dipartimento di Matematica e Fisica, Universit\`a Roma Tre, via della Vasca Navale 84 00146 Roma, Italy}
\author{Antonio Bianconi}
\affiliation{RICMASS Rome International Center for Materials Science, Superstripes Via dei Sabelli 119A, 00185 Roma, Italy}

\affiliation{Institute of Crystallography, CNR, via Salaria Km 29. 300, Roma I-00016, Italy}
\affiliation{National Research Nuclear University MEPhI (Moscow Engineering Physics Institute), 115409 Moscow, Russia}

\begin{abstract}
It is well known that the critical temperature of multi-gap superconducting 3D heterostructures at atomic limit (HAL) made of a superlattice of atomic layers with an electron spectrum made of several quantum subbands can be amplified by a shape resonance driven by the contact exchange interaction between different gaps. The $T_C$ amplification is achieved tuning the Fermi level near the singular nodal point at a Lifshitz transition for opening a neck. Recently high interest has been addressed to the breaking of inversion symmetry which leads to a linear-in-momentum spin-orbit induced spin splitting, universally referred to as Rashba spin-orbit coupling (RSOC) also in 3D layered metals. However the physics of multi-gap superconductivity near unconventional Lifshitz transitions in 3D HAL with RSOC, being in a non-BCS regime, is not known. The key result of this work getting the superconducting gaps by Bogoliubov theory and the 3D electron wave functions by solution of the Dirac equation is the feasibility of tuning multi-gap superconductivity by suitably matching the spin-orbit length with the 3D superlattice period.  It is found that the presence of the RSOC amplifies both the k dependent anisotropic gap function and the critical temperature when the Fermi energy is tuned near the circular nodal line. Our results suggest a method to effectively vary the effect of RSOC on macroscopic superconductor condensates via the tuning of the superlattice modulation parameter in a way potentially relevant for spintronics functionalities in several existing experimental platforms and tunable materials needed for quantum devices for quantum computing. 
\end{abstract}

 \date{\today}   

\maketitle

\section{Introduction}
It is known that the structure inversion asymmetry (SIA) which stems
from the inversion asymmetry of the confining potential in a 2D electron
gas induces a spin-orbit band splitting with states of different helicity
[\onlinecite{bychkov1984properties, rashba1960properties, ramaglia2003conductance,zhang2014fermi,perroni2007rashba,ramaglia2006ballistic,caprara2012intrinsic,bucheli2014phase,caprara2014inhomogeneous}]. Giant spin-orbit induced spin splitting
in the range 150--450 meV has been found in metal alloys [\onlinecite{ast2007giant}]
and transition-metal dichalcogenides [\onlinecite{sakano2012three}]. A three
dimensional Rashba spin splitting has been observed in PtBi$_{2}$,
BiTeX (X = Br, Cl, or I ) and GeTe which show dispersion along the
out-of-plane direction ($k_z$) [\onlinecite{ishizaka2011giant,liebmann2016giant,feng2019rashba}].
The realization of the three-dimensional Rashba-like spin splitting
[\onlinecite{brosco2017anisotropy}] in quantum materials and heterostructures potentially
unfolds numerous promising applications. Following the first theoretical
study of superconductivity [\onlinecite{gor2001superconducting}]  with spin-orbit
band splitting in a 2D metallic layer or at the surface of doped WOx
oxides, several theoretical works have studied the emergence of superconductivity
in the presence of spin-orbit coupling in a 2D metallic layer [\onlinecite{chaplik2006bound,cappelluti2007topological,vyasanakere2011bound,takei2012low,goldstein2015band,hutchinson2018enhancement,allami2019superfluid}]. 

Recently, experimental evidence that the strength of spin-orbit interaction
is correlated with quasi 2D superconductivity in the (111) LaAlO$_3$/
SrTiO$_3$ interface has been reported [\onlinecite{stornaiuolo2017signatures}] and confirmed
in several systems [\onlinecite{massarotti2020high,rout2017link,ptok2018superconducting,liu2020discovery}]. The spin polarized energy bands near a topological Lifshitz transition can be detected experimentally by ARPES spectroscopy as it has been observed in complex oxide heterostructure interface [\onlinecite{mori2019controlling}] and in layered cuprate perovskite superconductors [\onlinecite{gotlieb2018revealing}]. Today there is a high interest in the physics of quantum complex materials aimed at  the realization of mesoscopic quantum heterostructures
for novel superconducting Josephson junctions [\onlinecite{stornaiuolo2019high}-\onlinecite{caruso2020low}].

The theoretical studies of superconductivity coexisting with spin-orbit
coupling have been limited to a 2D superconducting layer and to a
single band metal [\onlinecite{gor2001superconducting,chaplik2006bound,cappelluti2007topological,vyasanakere2011bound,takei2012low,goldstein2015band,hutchinson2018enhancement,allami2019superfluid}],
while it is not known how superconductivity will arise in a 3D Rashba
system. Moreover, previous theoretical investigations have considered single-gap
superconductors while, in multi-band 3D superconductors,  multiple-gap superconductivity, in the clean
limit, need to be considered in the presence  of band spin splitting due
to spin-orbit coupling.

In fact, in multi-gap superconductivity, it is no longer possible to neglect
the key role of quantum configuration interaction between superconducting
gaps as, for example, the BEC-BCS crossover gap at Lifshitz transitions
near a band edge and other gaps in the BCS limit far from band edges
[\onlinecite{valletta1997electronic,bianconi1998superconductivity,bianconi2005feshbach,innocenti2010resonant,shanenko2012atypical,bianconi2014shape,jarlborg2016breakdown,mazziotti2017possible,cariglia2016shape}].
Finally, all theoretical approaches have been developed in the BCS
regime where the Fermi energy is much higher of both the spin-orbit
energy band splitting and the energy gap, while the most interesting
physics occurs in the regime where the Fermi energy is in the same
energy range as the superconducting energy gaps and the spin-orbit-splitting.

The main results of this work is the theoretical description of multi-gap superconductivity 
 [\onlinecite{valletta1997electronic,bianconi1998superconductivity,bianconi2005feshbach,innocenti2010resonant,innocenti2010shape,shanenko2012atypical,bianconi2014shape,jarlborg2016breakdown,mazziotti2017possible,cariglia2016shape,innocenti2010shape}] at the unconventional
Lifshitz transition [\onlinecite{volovik2018exotic}] in a 3D heterostructure at the atomic limit with a periodicity 
of few nanometers with tunable spin-orbit strength. 

We consider a 3D superlattice of metallic layers of thickness $L$ separated
by spacers of thickness $W$ and overall periodicity $d$. Our aim is to show that
the interplay between the Rashba spin-orbit coupling (RSOC) and superlattice
structure allows for a fine tuning of the critical temperature. To appreciate this point, consider
the energy splitting due to the RSOC and the corresponding difference
of the Fermi momenta of the two spin eigenstates. This difference
introduces a typical SOC length scale $l_{SOC}$, which may be compared
with the modulation of the superlattice $d$. In a bulk system $l_{SOC}$
can be compared only with the Fermi wavelength, which is typically
of the order of $0.1$ nm. In contrast in a superlattice, the modulation
is of the order of tens of nm, which matches the order of magnitude
of the RSOC. The RSOC energy is linear in the wave vector $\epsilon\sim\alpha k$,
with the constant $\alpha\sim 0.01$ eV nm. By defining $l_{SOC}=2\pi\hbar^{2}/\left(\alpha m\right)$,
$m$ being the electron mass, one estimates $l_{SOC}\sim10$ nm. As
a result the tuning of the RSOC may be achieved via the variation
of the modulation of the superlattice structure.

The layout of the paper is the following. In the next section, we introduce the model Hamiltonian of a 3D layered superconductor in the presence of RSOC.
In section III we study the normal phase paying special attention to the topology of the Fermi surface and to the associated features in the single-particle density of states (DOS). In section IV we turn our attention to the superconducting phase where we derive the superconducting gap equation and discuss its numerical solution in the multi-band case. Finally, in section V we state our conclusions.

\section{The Model}
The Hamiltonian of the system under study reads

\begin{equation}
H=H_{0}+H_{I},\label{eq:ham}
\end{equation}
where $H_{0}$ is the single-particle contribution, which includes
the RSOC

\begin{equation}
H_{0}=\frac{\mathbf{p_{\parallel}}^{2}}{2m}+\frac{p_{z}^{2}}{2m_{z}}+V\left(z\right)-i\alpha\left(\sigma_{x}\hbar\partial_{y}-\sigma_{y}\hbar\partial_{x}\right).\label{eq:h0}
\end{equation}
In the above equation, $\mathbf{p}=-i\hbar\nabla$ is the usual momentum
operator and $\mathbf{p_{\parallel}}$ its projection in the xy plane.
$V\left(z\right)=V\left(z+d\right)$ is the periodic potential modeling
the superlattice structure $V\left(z\right)=-V\left[\theta\left(z-d\right)-\theta\left(z-L\right)\right]$,
where $d=L+W$ and $V$ is a positive constant. The single-particle
Hamiltonian $H_{0}$ has solutions of the form
\begin{equation}
\psi_{n\mathbf{k}\lambda}\left(\mathbf{r}\right)=\varphi_{nk_{z}}\left(z\right)\frac{e^{i\mathbf{k}_{\parallel}\cdot\mathrm{r}_{\parallel}}}{\sqrt{\mathcal{A}}}\boldsymbol{\eta}_{\lambda}\left(\theta\right),\label{eq:wavefunction}
\end{equation}
where the wave vector components $\mathbf{k=}\left(k_{x},k_{y},k_{z}\right)\equiv\left(\mathbf{k}_{\parallel},k_{z}\right)$
label plane waves in the xy plane of area $\mathcal{A}$ and the Bloch
functions $\varphi_{nk_{z}}\left(z\right)$ along the z axis, $n$
being a subband index. The functions $\varphi_{nk_{z}}\left(z\right)$
and the corresponding eigenvalues are obtained by imposing the continuity
of the wave function and its first derivative at the discontinuity
points of the potential 
\begin{equation}
\varphi\left(z+d\right)=e^{ik_{z}d}\varphi\left(z\right),\;\varphi'\left(z+d\right)=e^{ik_{z}d}\varphi'\left(z\right),\label{eq:matching}
\end{equation}
where the phase factor is required by Bloch's theorem. Finally the
effect of the RSOC is encoded in the spinors
\begin{equation}
\boldsymbol{\eta}_{\lambda}\left(\theta\right)=\frac{1}{\sqrt{2}}\left(\begin{array}{c}
1\\
i\lambda e^{i\theta}
\end{array}\right),\quad\lambda=\pm1,\label{eq:spinors}
\end{equation}
where $\theta$ is the angle which defines the direction of the wave
vector in the plane $k_{x}=k_{\parallel}\cos\left(\theta\right)$, $k_{y}=k_{\parallel}\sin\left(\theta\right)$.
As a result the single-particle energies read

\begin{equation}
\epsilon_{n\mathbf{k}\lambda}=\varepsilon_{nk_{z}}+\frac{\hbar^{2}k_{\parallel}^{2}}{2m}+\lambda\alpha k_{\parallel}\equiv
\varepsilon_{nk_{z}}+\varepsilon_{\lambda k_{\parallel}}.\label{eq:energies}
\end{equation}
As for the second contribution to the Hamiltonian in Eq.(\ref{eq:ham}),
we adopt the standard contact interaction with a cut-off energy $\hbar\omega_{0}$
\begin{equation}
H_{I}=\frac{U_{0}}{2}\int\mathrm{d\mathbf{r}\varPsi_{\alpha}^{\dagger}\left(\mathbf{r}\right)\varPsi_{\beta}^{\dagger}\left(\mathbf{r}\right)\varPsi_{\beta}\left(\mathbf{r}\right)\varPsi_{\alpha}\left(\mathbf{r}\right),}\label{eq:interaction}
\end{equation}
where $\varPsi_{\alpha}\left(\mathbf{r}\right)$ is the annihilation fermion
field operator and summation over the repeated spin indices ($\alpha, \beta$)  is understood. 

Before considering the superconducting phase in section IV, it is useful to analyze first in the next section the effects of the RSOC in the normal phase and in particular on the density of states.
To this end, we first consider a simplified tight-binding model and then we turn our attention to the model defined in Eq.(\ref{eq:h0}), by confining to the two lowest  subbands for numerical reasons.

For the following discussion it is useful to introduce two dimensionless parameters: the Lifshitz parameter defined as

\begin{equation}
\eta=\frac{\mu-E_2}{\hbar\omega_0}
\label{eq:lif}
\end{equation} 

and the rescaled Lifshitz parameter

\begin{equation}
\eta_R=\frac{\mu-E_R}{\hbar\omega_0}, \ \ \   with \ \ \ E_R=E_2-\Delta E_{RSOC}
\label{eq:lifr}
\end{equation} 

where $ \mu $ is the chemical potential which at zero temperature coincides with the Fermi energy, $ E_2 $ is the band edge energy of the second subband in the absence of RSOC, $ \omega_0 $ is the cut- off and $ \Delta E_ {RSOC} $ is the energy shift due to the RSOC.

\section{The Normal Phase}
In the presence of a RSOC, the trend of the DOS can be understood by considering the evolution of the Fermi surface. In this context we will limit our analysis to a two-band system obtained by taking the two lowest subbands.

The starting point is the single-particle energy dispersion (\ref{eq:energies}), which we report here for the  sake of clarity:
\begin{equation}
\varepsilon_{n\textbf{k}\lambda}=\frac{k^2_\parallel}{2m}+\lambda \alpha k_\parallel +\varepsilon_{nk_z},
\label{eq:disp}
\end{equation} 
where for simplicity we adopt units such that $\hslash=1$. 

For both the first and second subband, the energy dispersion along the $z$ axis, which is numerically solved as shown below,  can be  fitted in terms of a tight-binding model. In particular, for odd n the agreement is obtained with a two-harmonic expansion, while for n even the agreement is obtained with a three-harmonic expansion. 
All this can be combined with the observation that, for the purpose of the subsequent discussion, we do not need to specify the precise form of the dispersion along the z-axis, but for the fact the $\varepsilon_{nk_z}$ increases (for n odd) or decreases (for n even)  monotonically between $k_z=0$ and $k_z=\pi/d$ and, furthermore, is an even function with respect to $k_z \longrightarrow -k_z$, for both even and odd $n$. 

Hence, in order to illustrate the key features of the DOS, we start our analysis with a simplified expression of $\varepsilon_{n k_z}$, namely:
\begin{equation}
\varepsilon_{1k_z}=t(1-cos(d\cdot k_z)), \ \ \ t=1\ \ {\rm and} \ 0<k_z<\pi/d
\label{eq:TB1}
\end{equation} 
for the first subband and
\begin{equation}
\varepsilon_{2k_z}=t(1+cos(d\cdot k_z)), \ \ \ t=1\ \ {\rm and} \ 0<k_z<\pi/d
\label{eq:TB2}
\end{equation} 
for the second subband.

To simplify the notation in the following discussion,  the parameters of the in-plane dispersion in Eq.(\ref{eq:disp}) are expressed in units such that  $2m=1$ and we define the spin-orbit typical momentum $k_0=m\alpha$.

For the sake of definiteness we assume that the minimum energy for the z axis is zero and the maximum is $\varepsilon_z$, i.e., $\varepsilon_{2n+1,0}=0$ and  $\varepsilon_{2n+1,\pi/d}=\varepsilon_z$ for the odd subbands, while we have $\varepsilon_{2n,0}=\varepsilon_z$ and  $\varepsilon_{2n,\pi/d}=0$ for the even subbands. Hence from (\ref{eq:disp}) we take the zero of the energy at the origin in the in-plane momentum space. Thus the dispersion along $z$ for the first and second subband will be equal to $\Delta E_{zn}=\varepsilon_z$.

\begin{figure}
		\includegraphics[scale=0.36]{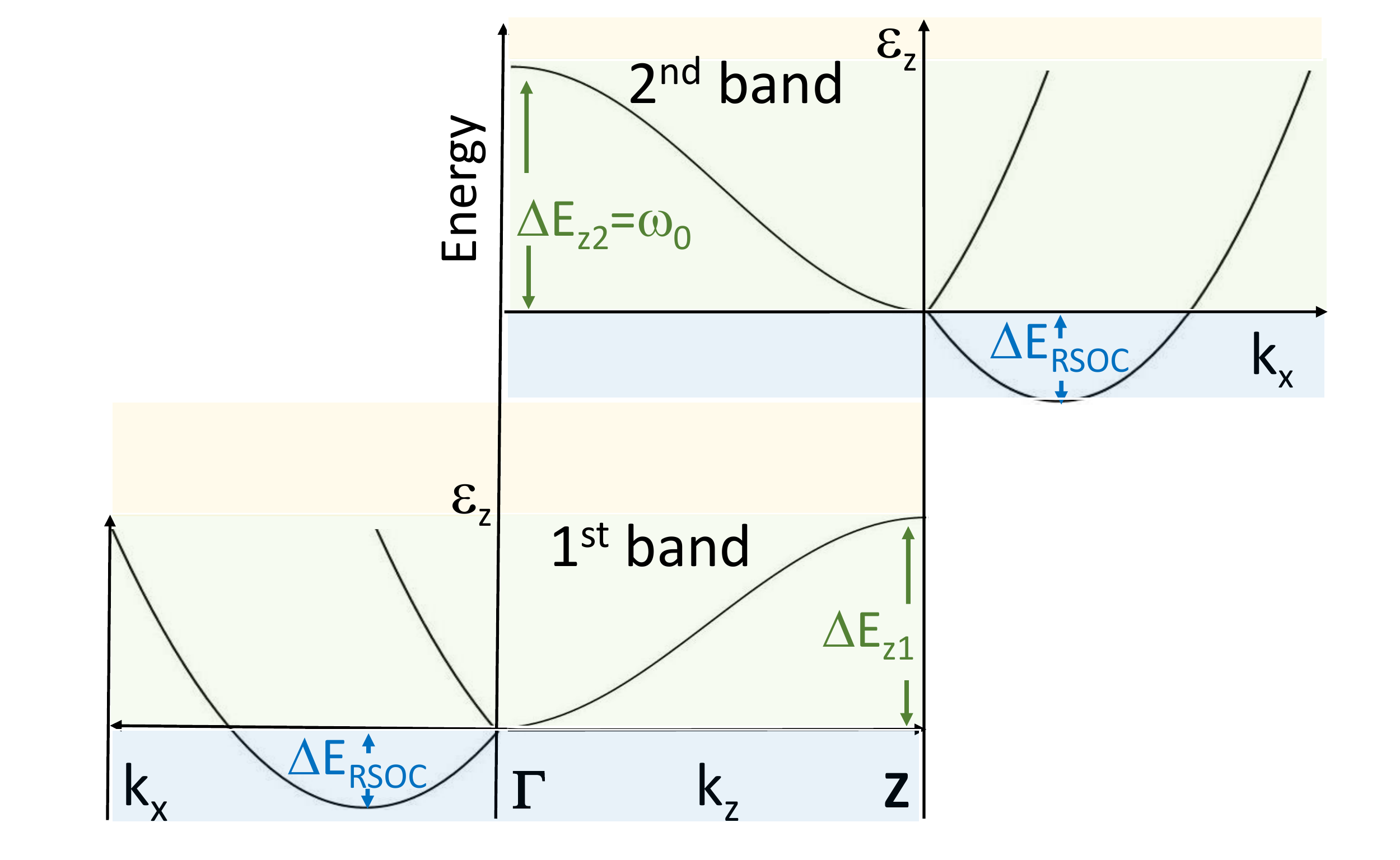}
	\caption{\foreignlanguage{english}{The dispersion along $ k_x $ is shown together with the dispersion along $ k_z $, in arbitrary units, both for n odd (bottom panel) and for n even (top panel). In this figure $ \Delta E_{zn} $ is the dispersion in the $ z $ direction equal to $ \varepsilon_z $ for both n even and odd (in the numerical model we will assume that $ \Delta E_{z2} $ is equal to the cut-off energy $ \omega_0 $). $ \Delta E_ {RSOC} = - k_0 ^ 2$ is, instead, the shift of the Dirac point (defined as the point at which the in-plane dispersions with opposite helicity meet) as a consequence of the dispersion along $ k_z $. The $\Gamma$ and $Z$ points are the center and the edge of the first Brillouin zone (IBZ).
						 We then indicate the three different regimes in which to study the system: I) $\varepsilon_z<\mu$ (light-yellow box); II) $0<\mu<\varepsilon_z$ (light-green box); III) $-k_0^2<\mu<0$ (light-blue box)}.}\label{fig1}
\end{figure}

The quasi particle energy (\ref{eq:disp}) has axial symmetry so that we may first study it in the $(k_\parallel, k_z)$-plane. From the isoenergetic curves in this plane one can obtain the isoenergetic surfaces by performing a rotation around the $k_z$ axis. At a given chemical potential $\mu$, from the expression of the quasiparticle energy we derive the values of $k_\parallel$ at fixed $k_z$ and helicity $\lambda$

\begin{equation}
k_\parallel (k_z,\lambda)=-\lambda k_0 \pm \sqrt{k_0^2+(\mu-\varepsilon_{n,k_z})},
\label{eq:Brancab}
\end{equation} 

from which we start our discussion. It is useful to distinguish three separate regimes for the Fermi energy: I) $\varepsilon_z<\mu$; II) $0<\mu<\varepsilon_z$; III) $-k_0^2<\mu<0$, where, in this simplified model, $\Delta E_{RSOC}=-k^2_0$ is the energy shift due to RSOC coupling (see the Fig.(\ref{fig1})).  

Let us examine them in detail.

\selectlanguage{american}%

\subsection{Regime I}
When selecting the sign in Eq.(\ref{eq:Brancab}) we must keep in mind that $k_\parallel \geq 0$. Let us start with the helicity $\lambda=1$. In this case for both even and odd n the only allowed sign is the positive one:

\begin{equation}
k_\parallel (k_z,1)=-k_0 + \sqrt{k_0^2+(\mu-\varepsilon_{n,k_z})}.
\label{eq:Brancap}
\end{equation} 
For $n$ odd at $k_z=0$ one has $k_\parallel (0,1)=-k_0+\sqrt{k^2_0+\mu}$, whereas at $k_z=\pi/d$ one has $k_\parallel (\pi/d,1)=-k_0+\sqrt{k^2_0+(\mu-\varepsilon_z)}$, so that $k_\parallel(\pi/d,1)<k_\parallel(0,1)$. Hence the isoenergetic curve, when rotated around the $k_z$ axis generates a corrugated cylinder wider in $ k_z=0 $ and narrower in $ k_z=\pm \pi/d$. For $n$ even we have a diametrically opposite situation, i.e., we still have a corrugated cylinder which, however, is narrower in $k_z=0$  and wider in $k_z=\pm \pi/d$ (Fig.(\ref{fig2})).

\begin{widetext}

\selectlanguage{american}%
\begin{figure}
	\includegraphics[scale=0.35]{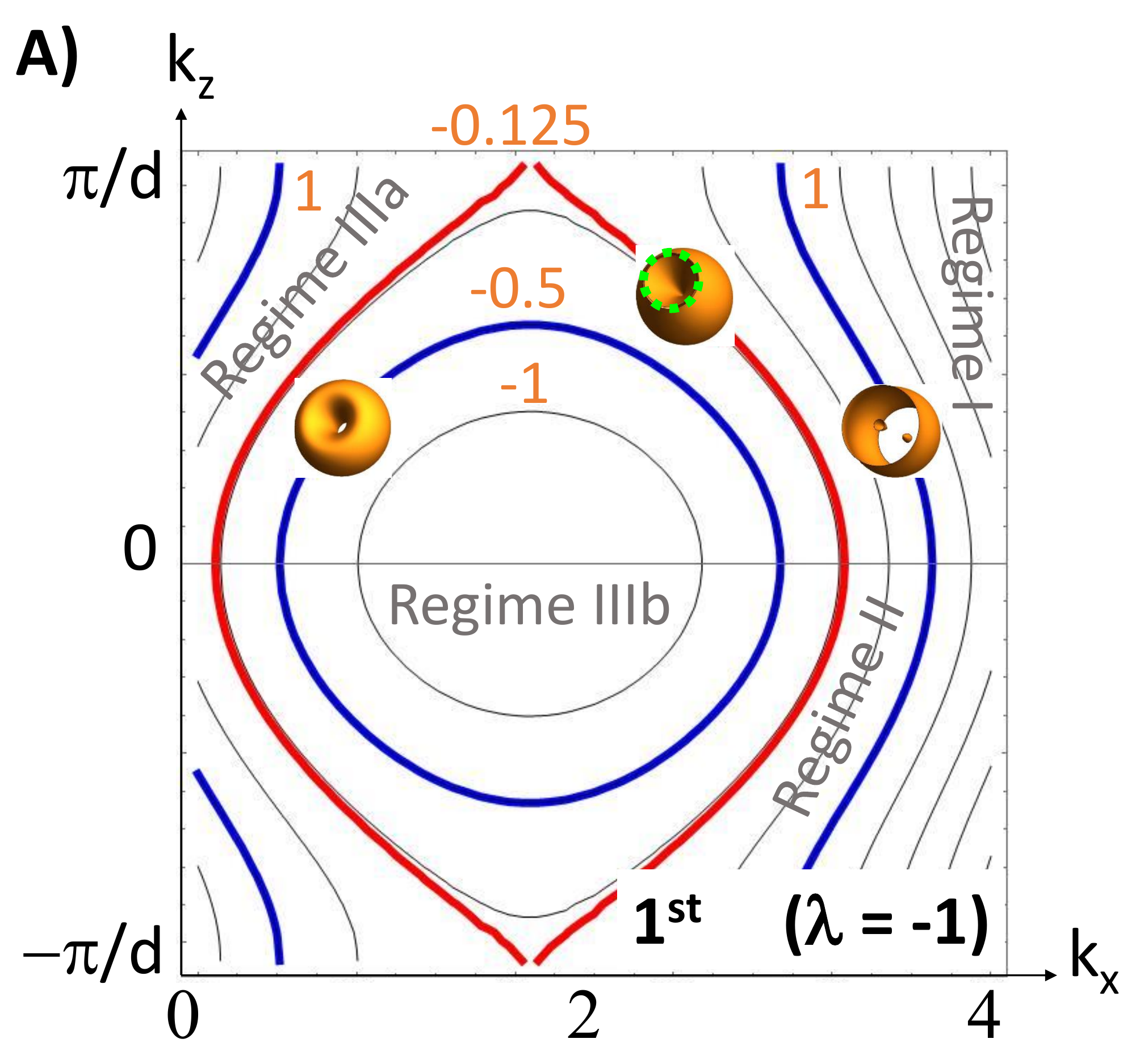}\quad\includegraphics[scale=0.35]{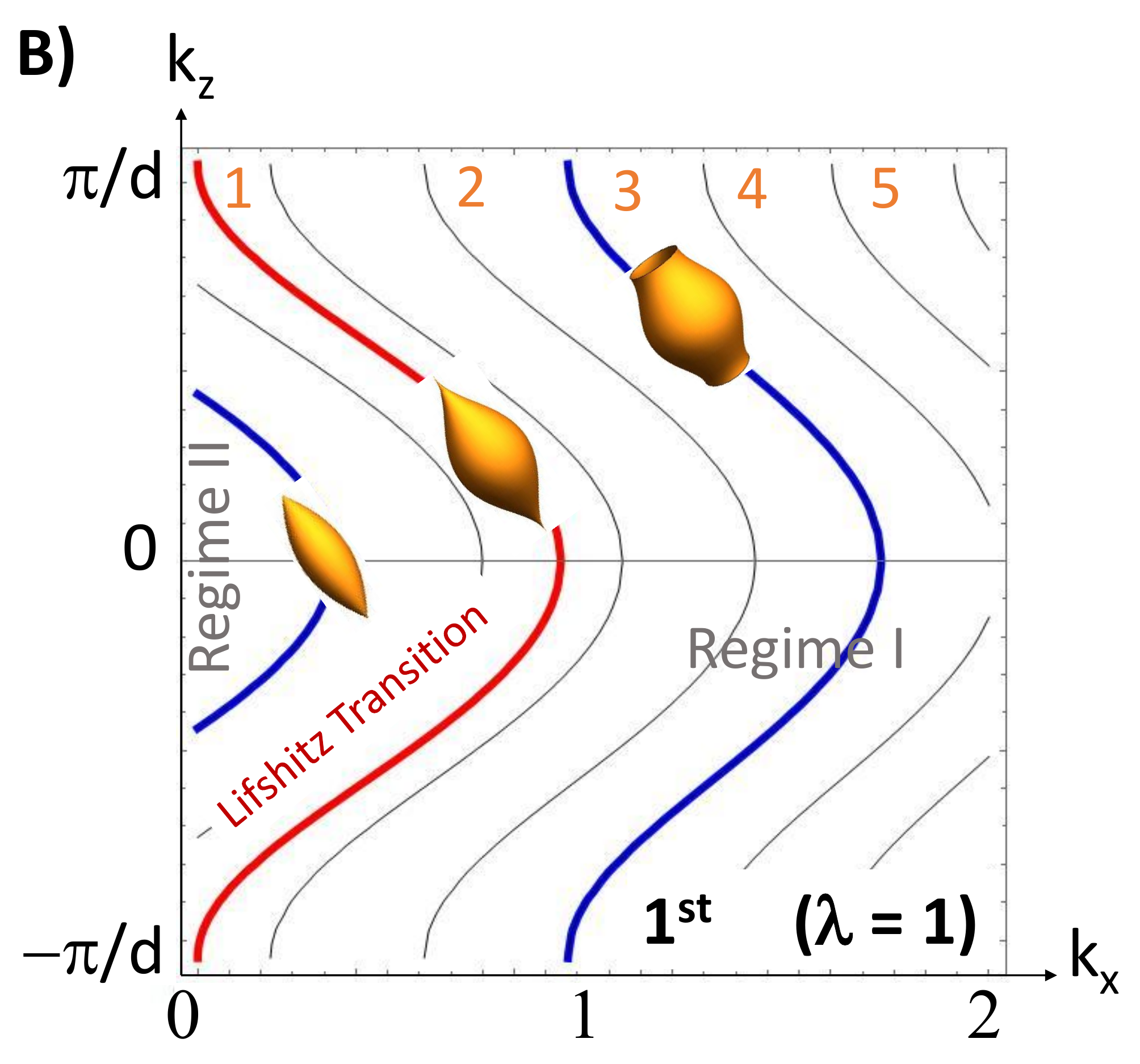}\quad 
	
	\includegraphics[scale=0.35]{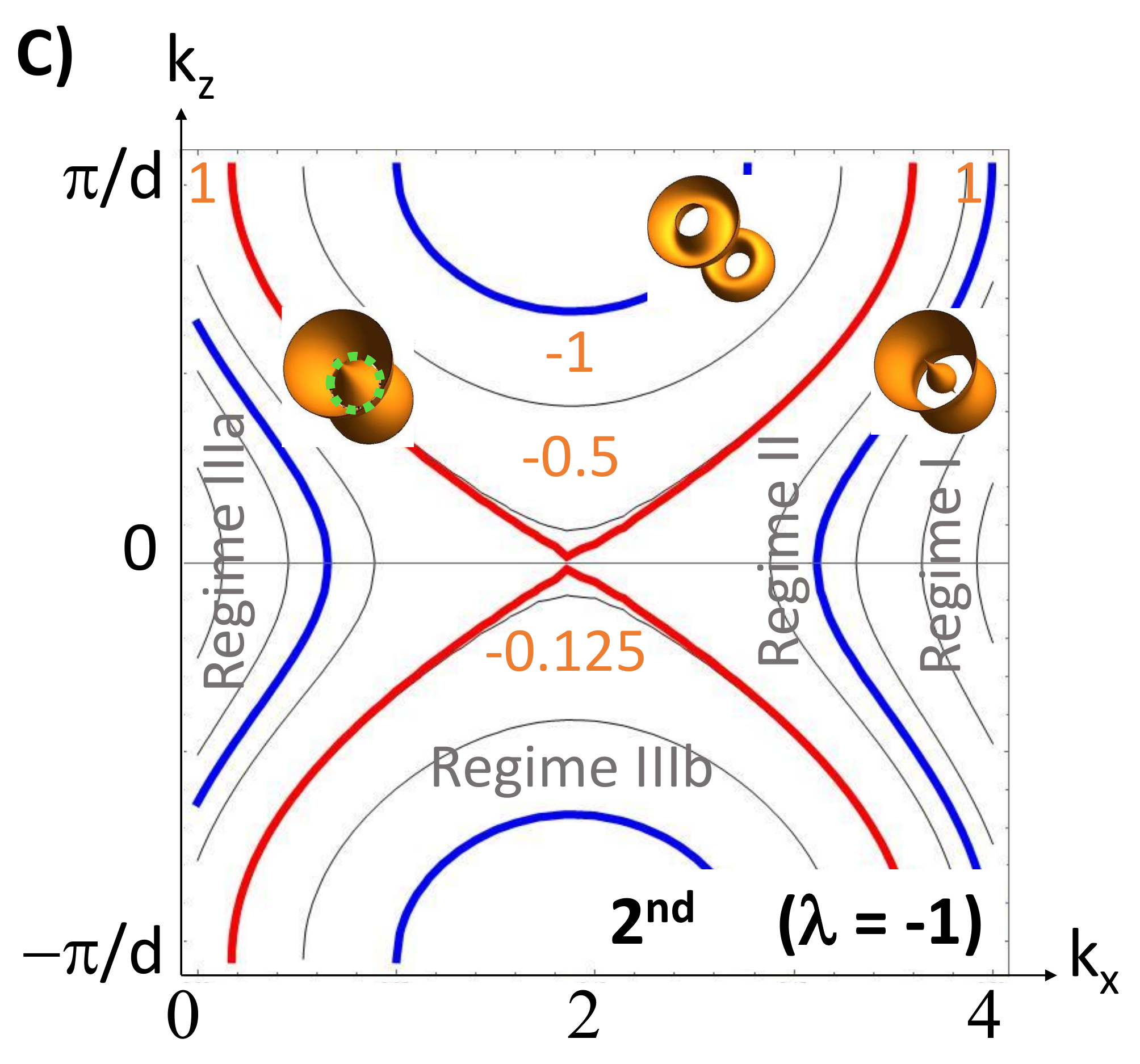}\quad\includegraphics[scale=0.35]{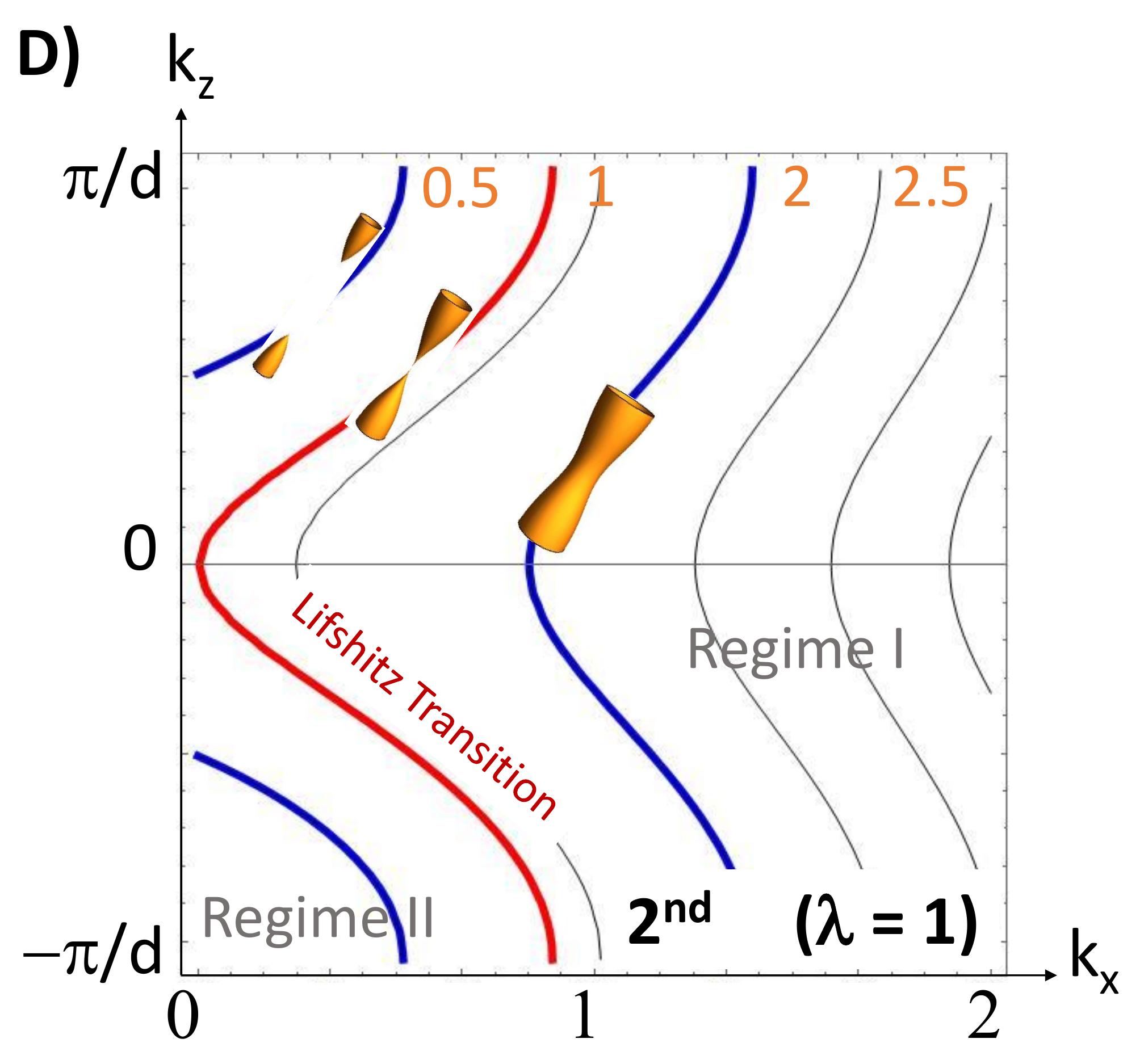}
	\caption{\foreignlanguage{english}{\textit{Top panel}: contour plots in the $(k_\parallel,k_z)$ plane for $\lambda=-1$ (left) and $\lambda=+1$ (right) for a single-harmonic tight-binding model for the first subband of Eq.(\ref{eq:TB1}). Parameters are $d=1$, $t = 1$ so that $\varepsilon_z=2$. The RSOC momentum $k_0 = 1.5$. On top of some of the isoenergetic curves are shown the corresponding Fermi surfaces. \textit{Bottom panel}:  contour plots as above for the second subband of Eq.(\ref{eq:TB2}). The orange numbers are the values of the Lifshitz parameter, defined in the Eq.(\ref{eq:lif}) of the different level curves, for the choice of the parameters made in this simplified model, while the dashed green curves on the 3D Fermi surfaces of panels A and C are represents the nodal line of singular points at an unusual van Hove singularity}.}\label{fig2}
\end{figure}

\end{widetext}

Let us consider next the case $\lambda=-1$.  Since $ \mu> \varepsilon_ {nk_z} $, the radicand is always greater than $ k_0 $, and, therefore, the only allowed sign is the positive one:
\begin{equation}
k_\parallel (k_z,-1)=k_0 + \sqrt{k_0^2+(\mu-\varepsilon_{n,k_z})} .
\label{eq:Brancam}
\end{equation} 
For $n$ odd at  $k_z=0$,  one has  $k_\parallel (0,-1)=k_0+\sqrt{k^2_0+\mu}$, whereas at $k_z=\pi/d$ one has $k_\parallel (\pi/d,-1)=k_0+\sqrt{k^2_0+(\mu-\varepsilon_z)}$. 
 Hence, also in this case, the isoenergetic curve, when rotated around the $k_z$ axis generates a corrugated cylinder, which is bigger than the previous one.

For $n$ even at  $k_z=0$  one has  $k_\parallel (0,-1)=k_0+\sqrt{k^2_0+(\mu-\varepsilon_z)}$, whereas at $k_z=\pi/d$ one has $k_\parallel (\pi/d,-1)=k_0+\sqrt{k^2_0+\mu}$. 
Hence, also in this case, the isoenergetic curve, when rotated around the $k_z$ axis generates a corrugated cylinder, with opposite curvature compared to the case of odd n (Fig.(\ref{fig2})).

\subsection{Regime II}
Let us begin again by considering first the helicity $\lambda=1$. Clearly the only sign allowed is the positive one. One notices that exactly at $\mu=\varepsilon_z$ one has $k_\parallel (\pi/d,1) = 0$,  which implies a Lifshitz transition for the Fermi surface. For the energies in this regime we see that not all the values of $k_z$ are allowed. The maximum $k_z=k^*_z$ is determined by the condition $k_\parallel(k^*_z,1)=0$ i.e. $k_0=\sqrt{k^2_0+(\mu-\varepsilon_{nk^*_z})}$. For odd $n$, the isoenergetic curve starts at a point $(0,k^*_z)$ on the $k_z$ axis and ends at a point $k_\parallel(0,1),0$ in the $k_\parallel$ axis. The Fermi surface has a fuse-like shape (Fig.(\ref{fig2})). 

For even $n$, the isoenergetic curve starts at a point $(0,k^*_z)$ on the $k_z$ axis and ends at a point  $k_\parallel(\pi/d,1),\pi/d$ on the  $k_\parallel$ axis.
In this case, for Fermi surfaces, we obtain half of a spindle that has the tip in $(0,0)$ and reaches the maximum diameter in $k_z=\pi/d$ (Fig.(\ref{fig2})).

In this regime the case for helicity $\lambda=-1$ is more complex. The positive sign is of course allowed. The branch with the positive sign starts at the point $(k_\parallel(\pi/d,-1),\pi/d)$ and ends at the point $(k_\parallel(0,-1),0)$ for odd $n$, while per even $n$ the positive sign starts at the point $(k_\parallel(0,-1),0)$ and ends at the point $(k_\parallel(\pi/d,-1),\pi/d)$. In both cases these curves generate corrugated cylinders by rotation around $ k_z $ (Fig.(\ref{fig2})). 
For this helicity there is also a possibility of the other branch with the negative sign:
 \begin{equation}
 k_\parallel (k_z,\lambda)=k_0 - \sqrt{k_0^2+(\mu-\varepsilon_{n,k_z})}.
 \label{eq:Brancamm}
 \end{equation} 
 However this branch is only allowed for a restricted range of $k_z$ values, i.e. $(k^*_z,\pi/d)$ which is the complementary range with respect to that allowed for the other helicity. Hence in this regime of energies the helicity $\lambda=1$ does not exist for the range $(k^*_z,\pi/d)$, when the helicity$\lambda=-1$ develops another branch exactly in this range. As a result the Fermi surface for the $\lambda=-1$ gets a apple-like shape with the poles pushed inwards. This is due to the fact that the points where the phase velocity vanishes are no longer isolated points, but due to  the rotation around $ k_z $ they form  circles with finite measure. 
 
  \subsection{Regime III} 
In this regime there is only the helicity $\lambda=-1$, which however has two branches:
 \begin{eqnarray}
k_\parallel (k_z,-1)&=&k_0 - \sqrt{k_0^2+(\mu-\varepsilon_{n,k_z})},
\label{eq:Brancammm}\\
k_\parallel (k_z,-1)&=&k_0 + \sqrt{k_0^2+(\mu-\varepsilon_{n,k_z})}.
\label{eq:Brancammp}
\end{eqnarray} 
If $k^2_0+\mu-\varepsilon_z=0$ then both branches start at the same point $(k_0,\pi/d)$ for odd n ($(k_0,0)$ for even n) and from there depart ending at the points $(k_0+\sqrt{k_0^2+\mu},0)$ and $(k_0 - \sqrt{k_0^2+\mu},0)$ for odd n ($(k_0+\sqrt{k_0^2+\mu},\pi/d)$ and $(k_0 - \sqrt{k_0^2+\mu},\pi/d)$ for even n) in the $k_\parallel$ axis, respectively. This is the case when the singularity in the phase velocity, which in the absence of RSOC is at the isolated point $(0,\pi/d)$ for odd n or  $(\pi/d,0)$ for even n, becomes a finite-measure manifold and develops a van Hove singularity in the DOS (Fig.(\ref{fig2})).
Hence we may distinguish two cases: IIIa) $0<k^2_0+\mu<\varepsilon_z$ and IIIb) $0<k^2_0+\mu>\varepsilon_z$. In the case IIIa) the argument of the square root is negative, hence the two branches start at a point $(k_0,k^{**}_z)$  with $k^{**}_z$ given by the condition $k^2_0+\mu=\varepsilon_{nk^{**}_z}$. Then the two branches end on the $k_\parallel$ axis. The Fermi surface generated by these curves has a torus-like shape. In regime IIIb) instead, the two branches remain disconnected from each other. The Fermi surface has an external and internal part and has a torus-like shape, with the toruses of neighboring zones touching each other (Fig.(\ref{fig2})).

Our aim is to evaluate the density of states (DOS) in order to compare it with the detailed calculations made with the more realistic periodic potential model. Therefore, we derive the analytical DOS expression for both helicity values, $ \lambda = \pm 1 $:

\begin{equation}
\begin{split}
N_- (\mu) &= \frac{1}{2\pi^2} \int_{-k_0}^\infty
dx(x + k_0) \\
&
\frac{\theta(\mu+k_0^2-x^2)\theta(x^2+2t-\mu-k_0^2)}{\sqrt{(\mu+k_0^2-x^2)(x^2+2t-\mu-k_0^2)}}
\end{split}
\label{eq:DOSMeno}
\end{equation}

\begin{equation}
\begin{split}
N_+ (\mu) &= \frac{1}{2\pi^2} \int_{k_0}^\infty  dx(x-k_0)\\
& \frac{\theta (\mu+k_0^2-x^2) \theta (x^2+2t-\mu-k_0^2)}{\sqrt{(\mu+k_0^2-x^2)(x^2 + 2t-\mu-k_0^2)}}
\end{split}
\label{eq:DOSPiu}
\end{equation}

where $x = k_\parallel\mp k_0$ for $\lambda = \mp 1$ and $\theta(x)$ is the Heaviside step function.

Let us analyze the integral defined in the Eq.(\ref{eq:DOSMeno}) and in the Eq.(\ref{eq:DOSPiu}). As a function of the variable $x$, the integrand has singularities at $x=\pm \sqrt{\mu+k_0^2}$ and $x= \pm \sqrt{\mu+k_0^2-2t}$. All singularities have index $-1/2$ and hence are integrable. When $\mu+k_0^2= 2t$, the denominator acquires a zero at the origin. In the absence of spin-orbit interaction, the $1/|x|$ behavior of the denominator is compensated by the numerator and the integral is finite. However, in the presence of spin-orbit interaction, there is a term proportional to $k_0$ in the numerator and a van Hove singularity develops. The singularity has a logarithmic behavior.

\begin{figure}
	\centering
	\includegraphics[scale=0.25]{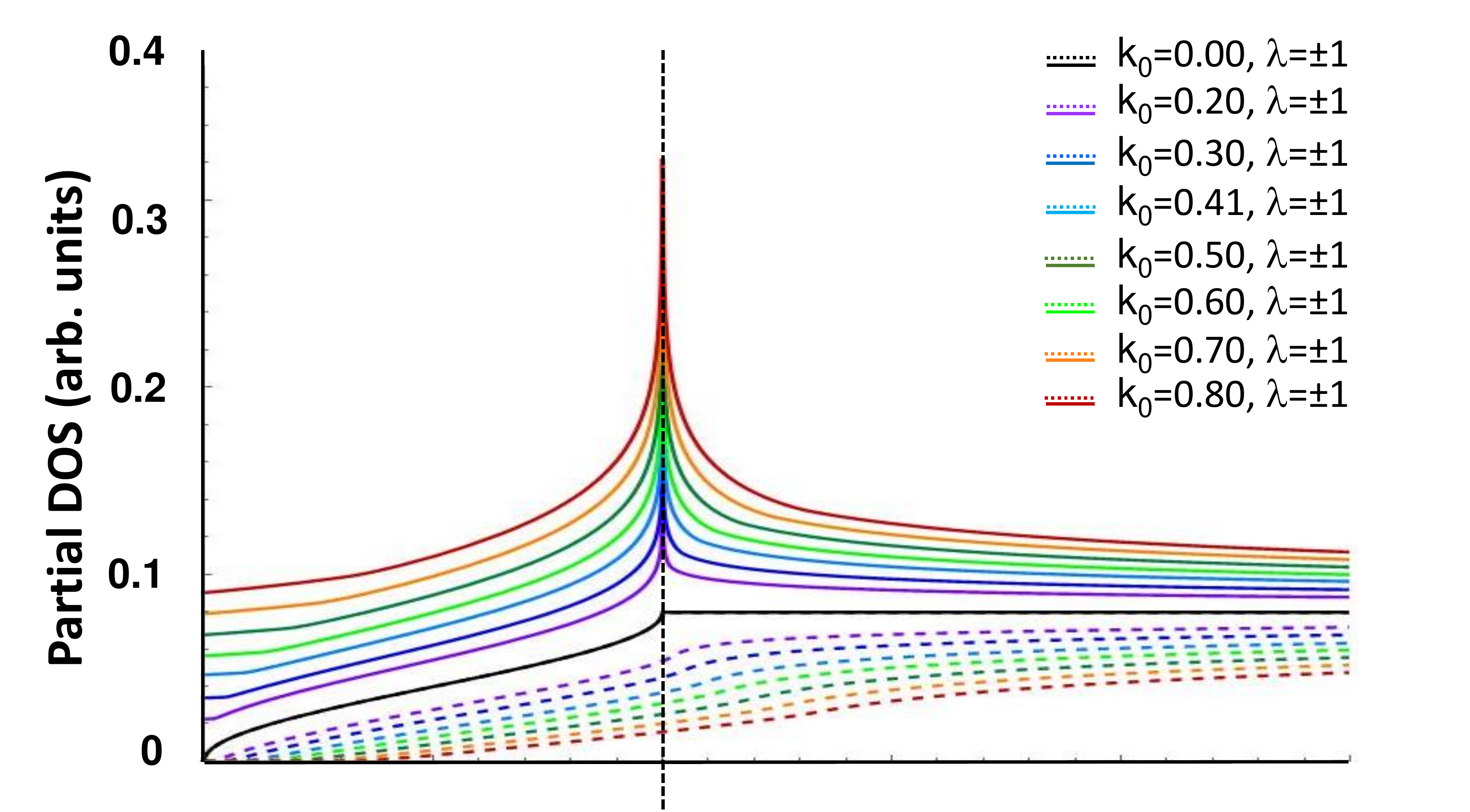}\quad \includegraphics[scale=0.25]{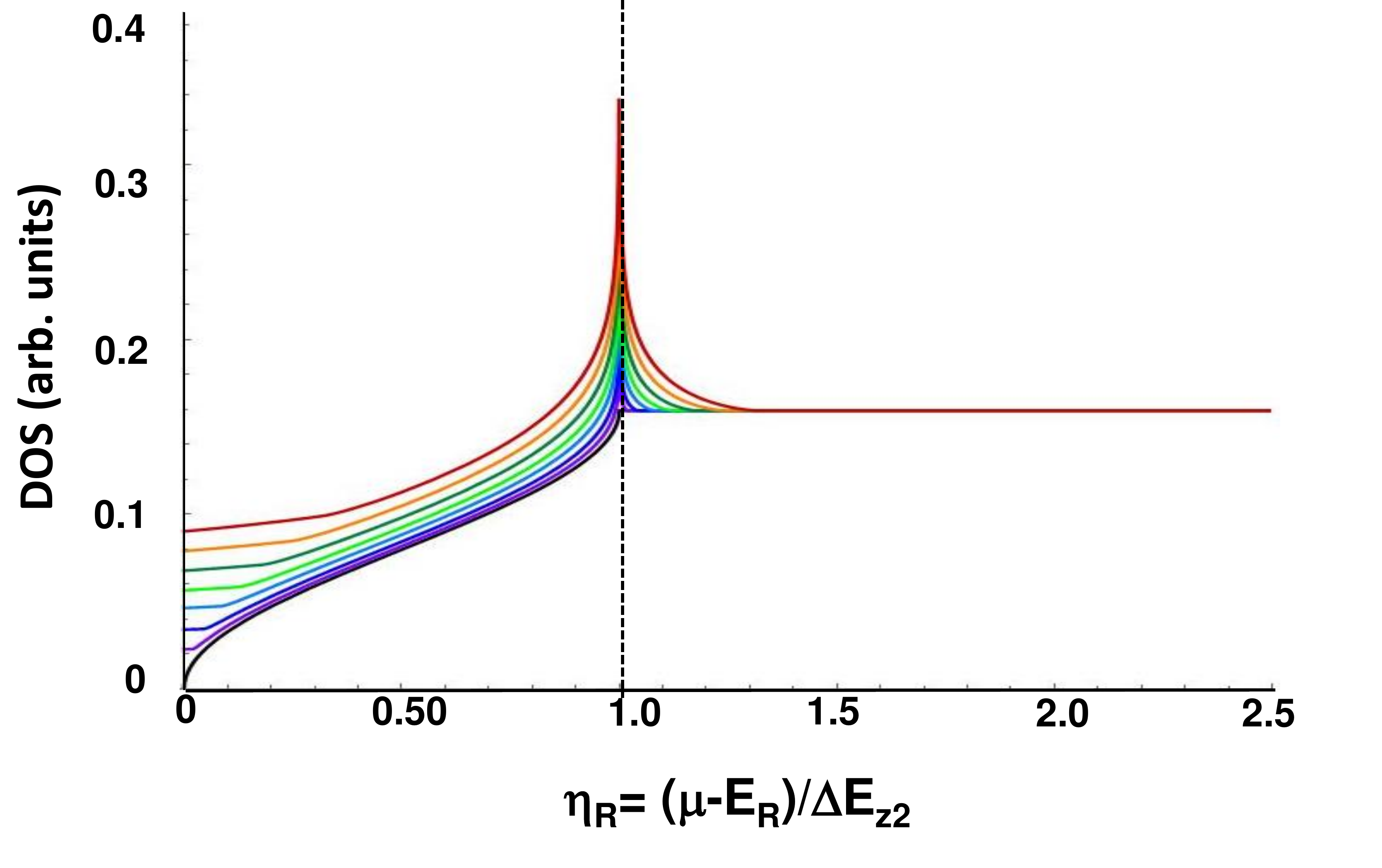} 
	\caption{\textit{Total and partial DOS as a function of $\eta_R$} (Eq.(\ref{eq:lifr})). \textit{Top panel}: the partial DOS $N_-$ (Eq.(\ref{eq:DOSMeno}))  and  $N_+$ (Eq.(\ref{eq:DOSPiu})) as function of the rescaled Lifshitz parameter $\eta_R$. The black curve is $k_0 = 0$. The other curves have increasing values of $k_0 = 0.1,\ 0.2,\ 0.3,\ 0.41,\ 0.5,\ 0.6, \ 0.7,\ 0.8$. Here $t = 1$. \textit{Bottom panel}: the total DOS $N_- + N_+$ given in Eqs.(\ref{eq:DOSMeno}), (\ref{eq:DOSPiu}) as function of rescaled Lifshitz parameter $\eta_R$. The black curve is $ k_0 = 0$. The other curves have increasing values of $k_0 = 0.1,\ 0.2,\ 0.3,\ 0.41,\ 0.5,\ 0.6, \ 0.7,\ 0.8$. Here $t = 1$.	
	}
	\label{fig3}
\end{figure}

The DOS expression can be computed with Mathematica by using the built-in Heaviside function and numerical integration command. In Fig.(\ref{fig3}) are reported the plots of $N_-$, $N_+$ (partial DOS), and  $N_-+N_+$ (total DOS), respectively for four values of $k_0 = 0.1,\ 0.2, \ 0.3, \ 0.41, \ 0.5, \ 0.6, \ 0.7, \ 0.8$ and $t = 1$. The partial and the total DOS are reported as a function of the rescaled Lifshitz parameter defined in the Eq.(\ref{eq:lifr}) where, in this case, $E_2=0$, $\Delta E_{RSOC}=-k^2_0$ and $\omega_0=\Delta E_{z2}=\varepsilon_z=2t$.

 The black curve corresponds to the case when there is no spin-orbit present.  Clearly, the value $\eta_R = 1$ (in units of $t$) marks the point of the band edge for the dispersion along the $z$ axis. The spin-orbit interaction develops a van Hove singularity exactly at this point. This behavior, as we will see below,  appears in agreement with the more realistic model. This point, $\eta_R = 2t/\Delta E_{z2}$, corresponds to the singularity in the two-dimensional Rashba model at the bottom of the lower band with helicity $\lambda =-1$. In the 3D case the singularity appears at the edge of the band due to the motion along $z$.

Fig.(\ref{fig3}), also, shows that as $ k_0 $ increases, $ N _ - $ increases while $ N _+ $ decreases, in the sum this involves a change only in the proximity of the van Hove singularity. More precisely, while at the Lifshitz transition the partial densities combine to yield a strong change in the DOS, at high energies they compensate,  so that the total DOS coincides with the total DOS in the absence of RSOC. This means that in the high-energy limit the parameters of the normal phase and, as we will see below, of the superconducting phase do not depend on $ k_0 $, in accordance with the work of Gorkov and Rashba [\onlinecite{gor2001superconducting}].

\subsection{Numerical results for the full model}

After the analysis of the simplified tight-binding model, we study the properties of the normal phase starting from the solution of the model of Eq.(\ref{eq:h0}) obtained numerically.

For the numerical solution of the normal phase  the chosen parameters are:
the barrier $V=0.5$ eV, the thicknesses of the metallic and insulating layers $L=23$ \AA, $W=7$ \AA, respectively, with total periodicity $d=30$ \AA , the effective masses $m=m_{z}=m_{e}$ , the cut-off energy $\hbar\omega_{0}=30\ meV$ and the coupling constant $g=0.4$. 

According to the works [\onlinecite{ramaglia2003conductance}, \onlinecite{ramaglia2006ballistic}], we express the Rashba coupling constant in the following form:
 \begin{equation}
\alpha =  2\frac{\hslash^2}{2m} \frac{2\pi}{d} \alpha_{SO},
\label{eq:alpha}
\end{equation} 
where $\alpha_{SO}$ is a dimensionless parameter which describes the strenght of the Rashba momentum in units of the inverse lattice spacing along the $z$ direction.

	Similarly to what has been done for the tight-binding model, we carry out the analysis of the evolution of Fermi surfaces for $\varepsilon_{2k_z} $ obtained numerically by distinguishing for each of the listed regimes three distinct cases: $\Delta E_{z2}\gtreqqless \Delta E_{RSOC}$, where $ \Delta E_{z2} = \varepsilon_ {2k_z, max} - \varepsilon_ {2k_z, min}$ is the bandwidth of the dispersion along the axis of confinement $z$ in the presence of a potential of the form Kronig-Penney, while in this case $\Delta E_{RSOC}=E_0 = -(m\alpha^2)/(2\hbar^2)$ is  the energy shift due to the RSOC. The model parameters are chosen so that $\Delta E_{z2}$ is of the same order of magnitude as the cut-off energy. 
	
	 This study  concerns a two-band system, where the first subband has a s-symmetry, while the second one has p-symmetry.  The results are shown in the Fig.(\ref{fig4}): in the panels A, B, C and D we plot the isoenergetic curves in the $(k_x, k_z)$-\textit{plane} for $\lambda=\pm 1$ and for the first and the second subbands versus the Lifshitz parameter, $\eta$, Eq.({\ref{eq:lif}}) where $ E_{2}=163.64\ meV $ is the band edge of the second subband in the absence of RSOC and $ \omega_0 $ is the cut-off energy for $\hslash = 1 $. In this figure we also report the evolution of Fermi surfaces for three distinct values of the parameter $ \eta $. The analysis is made for  $\alpha_{SO} = 0.41$ value for which the condition $\Delta E_{RSOC}=\Delta E_{z2}=\omega_0$  is verified. 
	In the case of the second subband for an energy value close to the van Hove singularity ($ \eta_L $) we take into account that in the presence of RSOC the spinor (Eq.(\ref{eq:spinors})), and the gap (Eq.(\ref{gapform})), depend on a phase factor $ e^{i \vartheta} $ and the removal of the spin degeneration splits the dispersion in two bands with opposite helicity. To take this into account we plot the FS at the Lifshitz transition with a color that varies with $ \vartheta $ in panels (c) and (d) of Fig. \ref{fig4}. In particular, for $ \lambda = 1 $ it varies from red to purple, while for $ \lambda = -1 $ it varies from purple to red.

\begin{widetext}
	
\selectlanguage{american}%
\begin{figure}
	\includegraphics[scale=0.35]{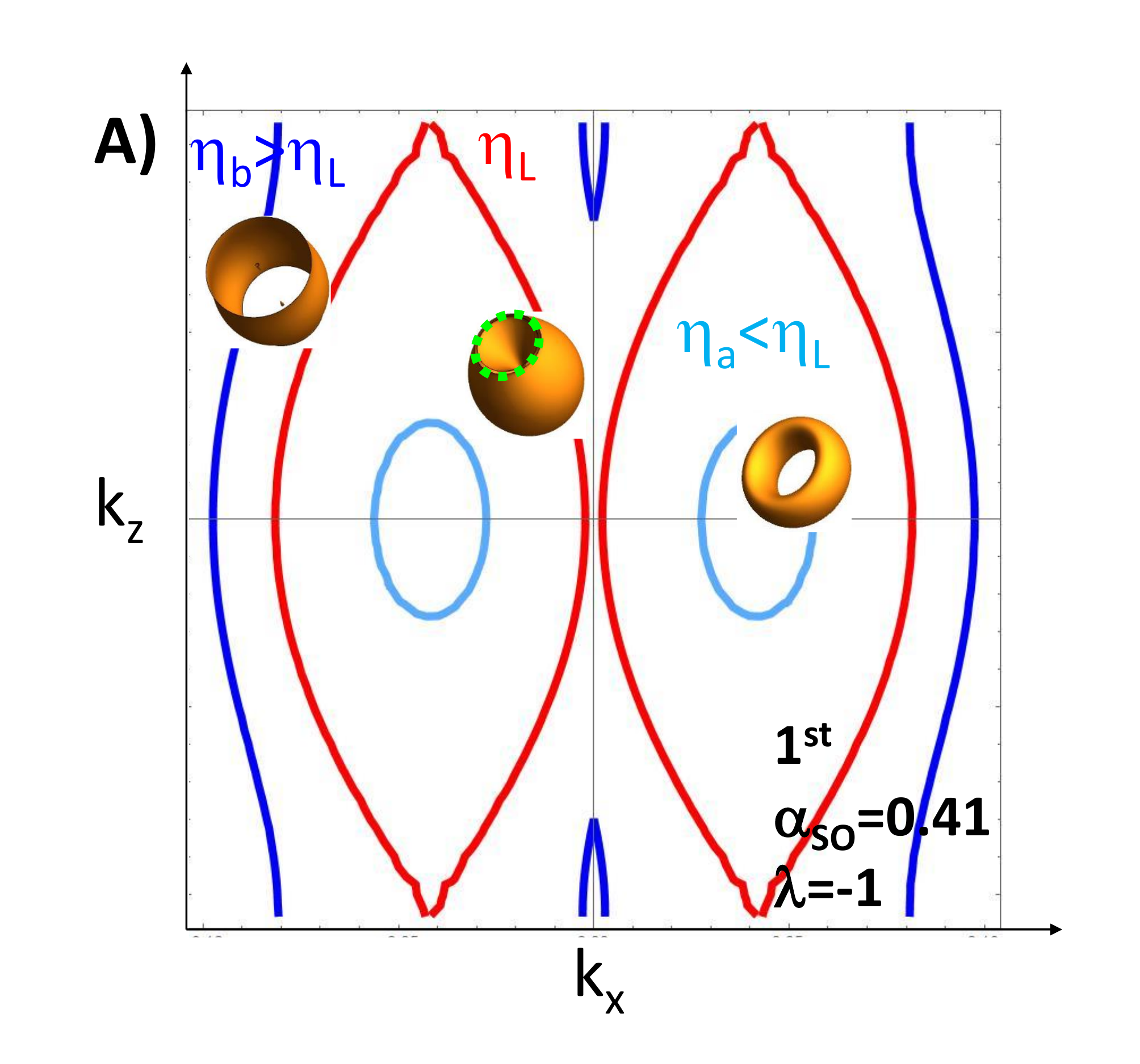}\includegraphics[scale=0.35]{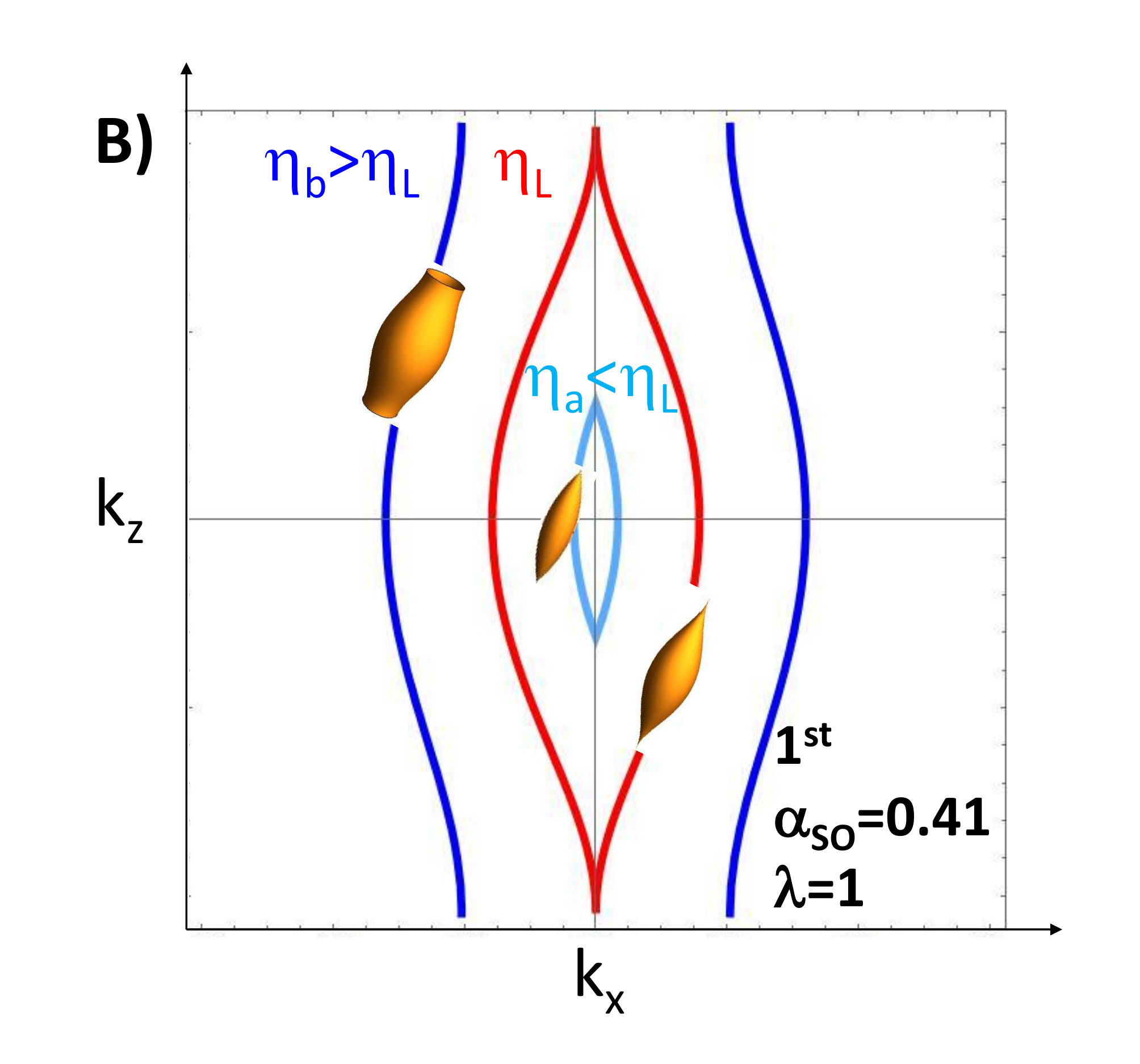} 
	\includegraphics[scale=0.35]{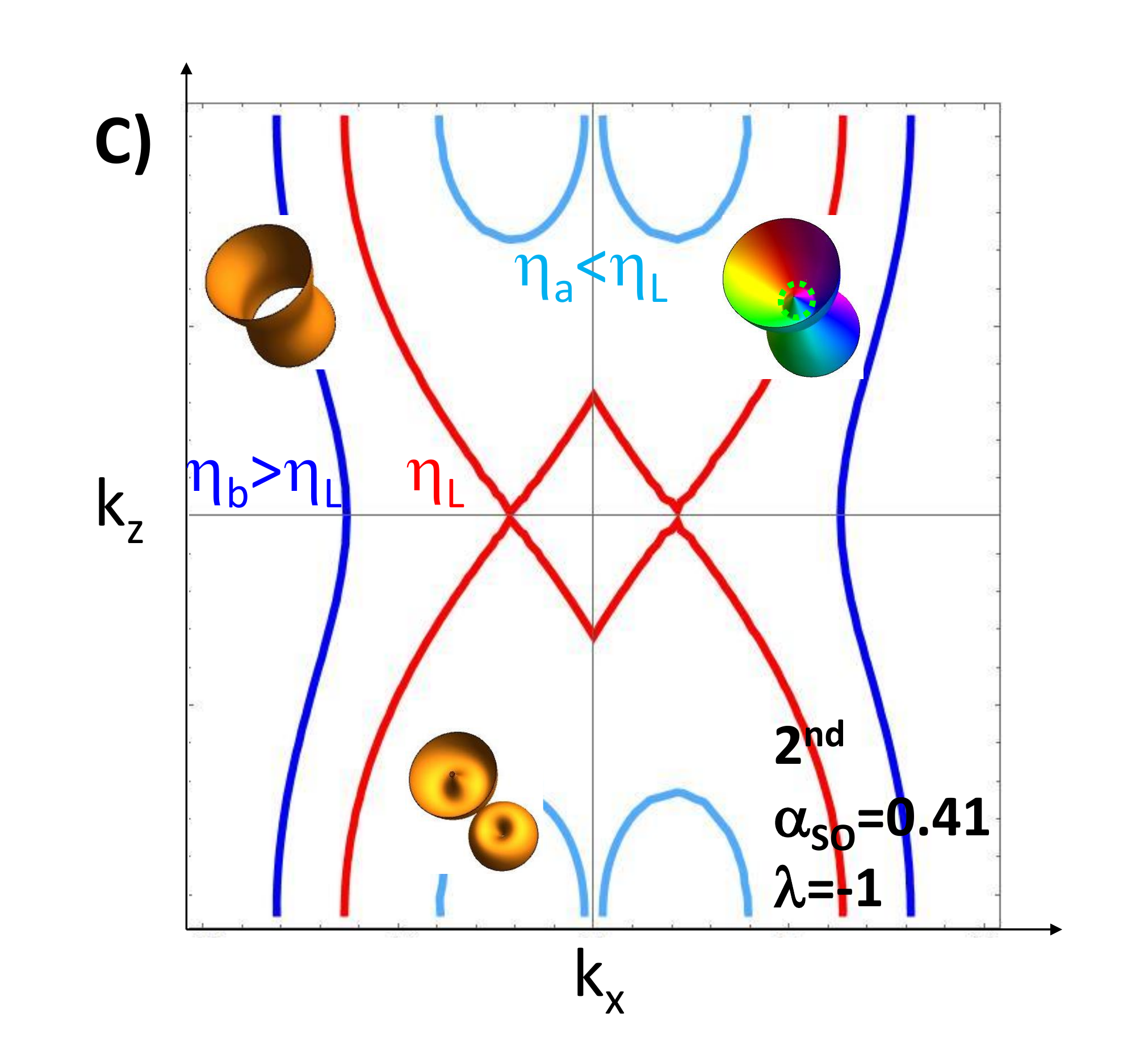}\includegraphics[scale=0.35]{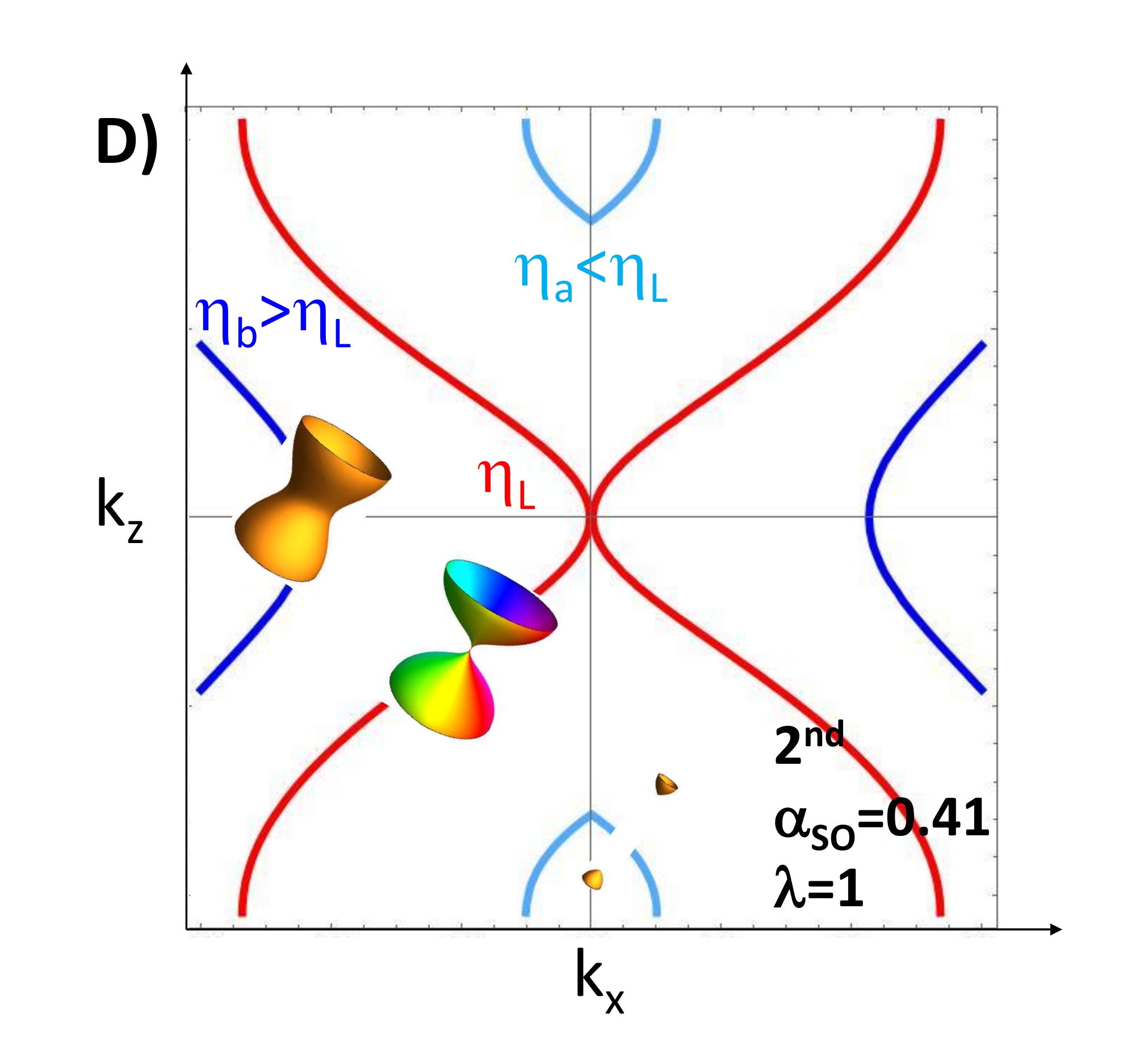}
				\caption{\foreignlanguage{english}{\textit{Isoenergetic curves, Fermi surfaces for the first and second subbands at three different values of Lifshitz parameter}.  The \textit{top panels} show the case of $ \lambda = 1 $ (left)  and  $ \lambda = -1 $ (right) for the first subband and for $ \Delta E_ {RSOC} = \Delta E_{2z} $ ($\alpha_{SO}=0.41$). The \textit{bottom panels }show the same analysis carried out for the second subband. The DOS maximum is observed at $ \eta_L $ where the system develops a van Hove singularity. In this case the Fermi surface develops a nodal line highlighted in panels A and C with a dashed green line. For $\lambda=1$, both for the first and for the second subband, $\eta_L$ is independently of the value of $\alpha_{SO}$ and it's equal to $-3.9\ meV$ and $0.92\ meV$, respectively. While for $\lambda=-1$, $\eta_L$ changes with $\alpha_{SO}$ (as underlined in Fig.(\ref{fig5})), for the first subband $\eta_L=-4.6\ meV$, for the second $\eta_L=-0.11\ meV$. In the panel C) and D), for $\eta=\eta_L$ we highlight the phase factor with the colors of the rainbow and the nodal line (white dashed curve) of the singular points.}}\label{fig4}
\end{figure}

\end{widetext}

We highlight the three regimes analyzed previously and a change in symmetry in passing from the first to the second subband. Such a change, for the first subband,  occurs at the point $ \Gamma $, origin of the first Brillouin zone (IBZ), while, in the second subband, it occurs at the point $ Z $, edge  of the IBZ in the $z$ direction. As the Rashba coupling changes ($\Delta E_z\gtreqqless \Delta E_{RSOC}$), only a flattening of the contour lines and Fermi surfaces is observed to the left and a shift to the right of the singular points. The latter are the points where the phase velocity vanishes and which generate, for rotation around the $ k_z $ axis,  circles whose radius increases with $\alpha_{SO}$. The energy in which this van Hove unusual singularity occurs is indicated in the figure with $ \eta_L $ and in the literature it is called neck opening energy. We can note that for the bands with positive helicity $ \eta_L $ is independent of the value of $ \alpha_ {SO} $, while for the bands with negative helicity it varies as the RSOC varies.

This behavior is confirmed by Fig.(\ref {fig5}) where we plot the partial DOS for $ \alpha_ {SO} = 0.20 $ ($\Delta E_{RSOC}> \Delta E_ {z2} $), $ \alpha_ {SO} = 0.41 $ ($\Delta E_ {RSOC}> \Delta E_ {z2} $) and $ \alpha_ {SO} = 0.50 $ ($\Delta E_ {RSOC}> \Delta E_ {z2} $). In  this figure $ \eta_L $ coincides with the value for which the partial DOS relative to $ \lambda = -1 $ for the first and second subband has a maximum. As you can see, as $ \alpha_ {SO} $ increases, the $ \eta_L $ parameter decreases while the value of the partial DOS peak $ \lambda = -1 $ increases. In particular, in the case of the first subband for $ \lambda = -1 $ $\eta_L=-5.2$ for $\alpha_{SO}=0.50$, $\eta_L=-4.6$ for $\alpha_{SO}=0.41$ and $\eta_L=-4.0$ for $\alpha_{SO}=0.20$. Indeed, in the case of the second subband for $ \lambda = -1 $ $\eta_L=-0.37$ for $\alpha_{SO}=0.50$, $\eta_L=0.11$ for $\alpha_{SO}=0.41$ and $\eta_L=0.78$ for $\alpha_{SO}=0.20$. In the right panel of Fig.(\ref {fig5}) we report the projection of the Fermi surfaces in the plane $( k_x, \ k_y )$ at the point $ Z $ of the IBZ for $ \eta = 3 $, where we have highlighted the two possible values of helicity with different colors (light blue for $ \lambda = -1 $ and orange for $ \lambda = 1 $).

\begin{widetext}

\selectlanguage{american}%
\begin{figure}
	\includegraphics[scale=0.5]{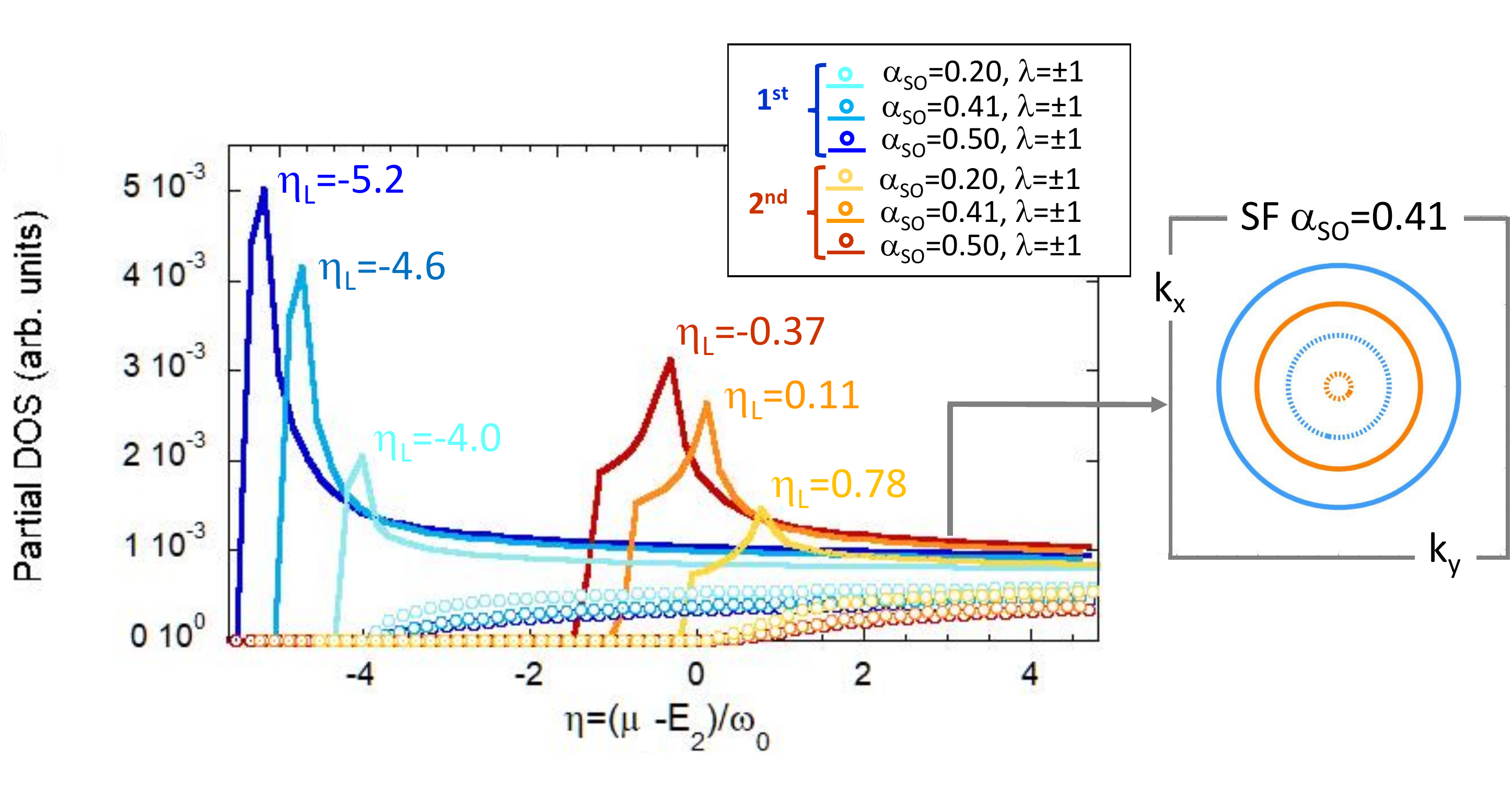}
	\caption{\foreignlanguage{english}{\textit{Partial density of the states vs $ \eta $ and projection of the FS in the plane $( k_x, \ k_y )$ at  $k_z= Z $}.  In the \textit{left panel} the continuous curves represent the partial DOS for $ \lambda = -1 $, the dots those for $ \lambda = 1 $, the shades of blue refer to the first subband for three different values of $ \alpha_ {SO } = 0.20, \ 0.41, \ 0.50 $, while the shades of red to the second subband. The figure shows that the value of $ \eta_L $, for which the FS have a nodal line (Fig.(\ref {fig4})), decreases as $ \alpha_ {SO} $ increases by an amount equal to $ E_0 / \omega_0 $, while the peak of the partial DOS $ \lambda = + 1 $ increases. In the \textit{right panel} the light blue  curve represents the projection of the FS relative to $ \lambda = -1 $ in the plane for the first subband, the light blue dashed curve is relative to $ \lambda = 1 $. Similarly, the orange curves refer to the second subband. This panel is built for $ \alpha_ {SO} = 0.41 $ and $ \eta = 3 $.}}\label{fig5}
\end{figure}

\end{widetext}

\begin{figure}
	\includegraphics[scale=1.55]{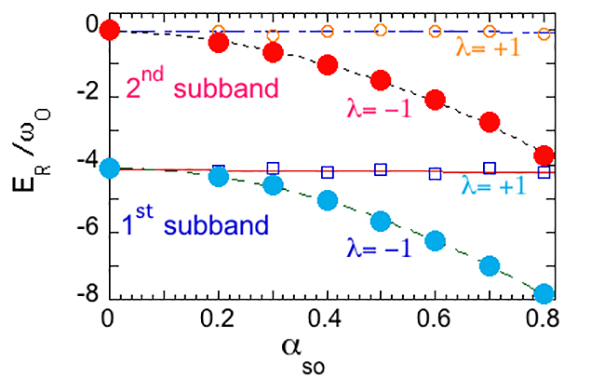}
	\caption{\foreignlanguage{english}{\textit{Normalized band-edge energy as a function of the RSOC constant}. The orange empty circles represent the band edge energy \textit{vs} $ \alpha_{SO} $ for the second subband and $ \lambda = 1 $, while the blue empty squares are relative to the first subband at the same helicity value. The red dots, on the other hand, represent the band edge energy for the second subband at $ \lambda = -1 $, the light blue dots refer to the first subband for the same helicity value. As noted earlier, a Rashba shift can only be observed for negative helicity bands.}}\label{fig6}
\end{figure}

What can be seen from the graphs in the Fig.(\ref{fig5}) is a peak in the partial DOS, and then in the total DOS, corresponding to an energy value equal to $ E_0/\omega_0 = - (m \alpha^ 2)/(2\omega_0) $ which, as the coupling constant Rashba increases,  it increases and shifts to gradually smaller Lifshitz parameter values, and the shift involves only the negative helicity bands. This is underlined in the Fig.(\ref{fig6}), in which we have reported the normalized band-edge energy, $E_R$ in the Eq.(\ref{eq:lifr}), for the first and the second subband for the two distinct helicity values as a function of the RSOC constant.

Finally,  note that, for the values of the normalized band-edge energy, from this point on, we report the results in terms of the rescaled Lifshitz parameter (Eq.(\ref{eq:lifr})).

\begin{figure}
	\includegraphics[scale=0.65]{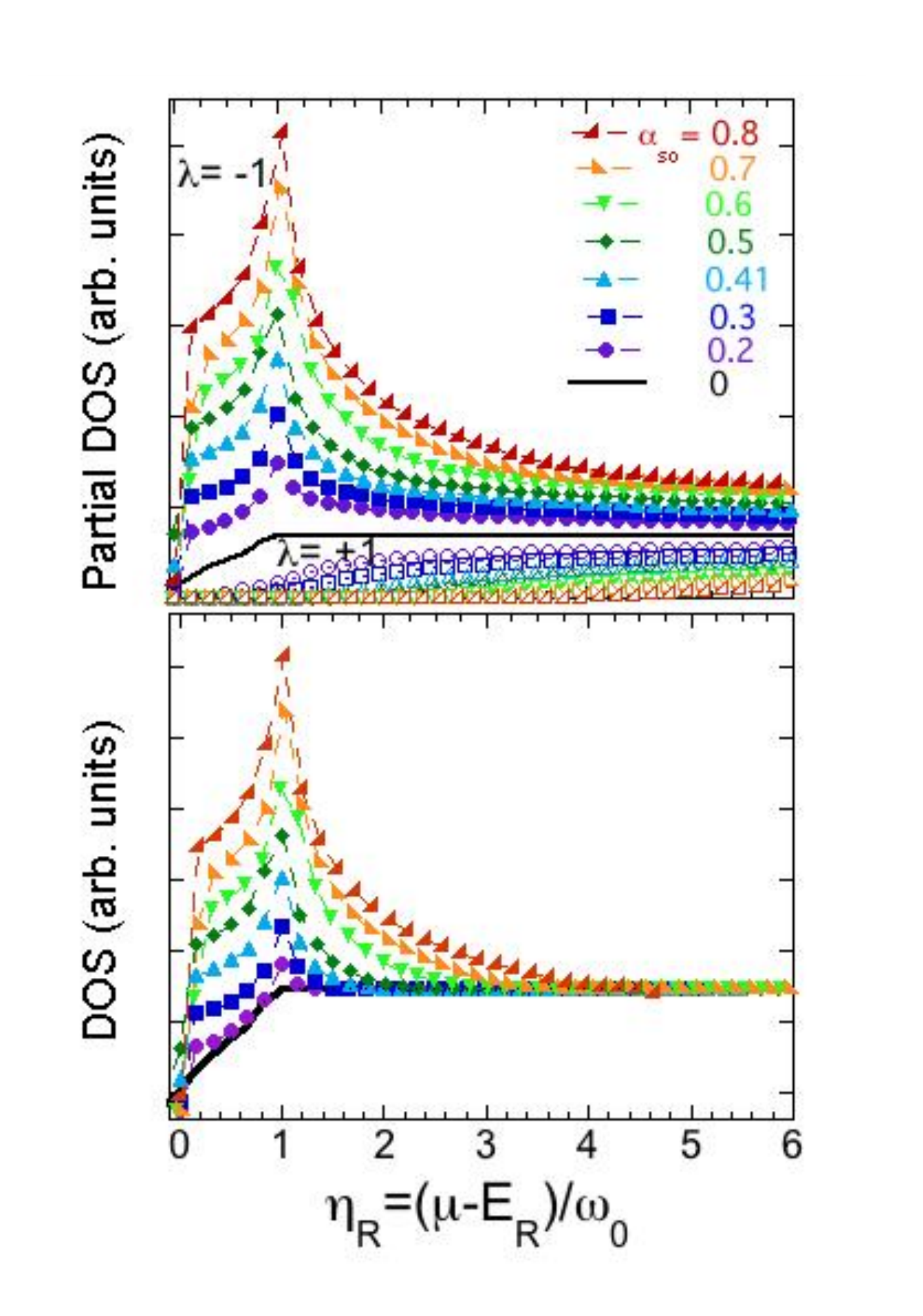}
	\caption{\foreignlanguage{english}{\textit{The DOS for the first and the second subband as the $\alpha_{SO}$ changes}. In particular, we have chosen values for this parameter between $0.2$ and $0.8$  in order to reproduce the three cases previously discussed $\Delta E_z\gtreqless \Delta E_{RSOC}$ and compare them with the case of no RSOC, $\alpha_{SO}=0$.
			What is observed is that as the RSOC increases, the DOS peak becomes more pronounced and shifts to gradually smaller energies, since the peak occurs at $ E_0/\omega_0$.}}\label{fig7}
\end{figure}

In the normal phase for a two-band system we plot the  total density of the states (DOS) and the partial DOS as the Rashba coupling changes ($\Delta E_z\gtreqqless \Delta E_{RSOC}$) and compare it with the case without RSOC (Fig.(\ref{fig7})). In Fig.(\ref{fig7}) the DOS is plotted versus the rescaled Lifshitz parameter, $\eta_R$ (Eq.(\ref{eq:lifr})) for different values of $\alpha_{SO}$. 

In the case of positive helicity the DOS trend is that of a sloped step, very similar to the trend observed in the absence of RSOC, while in the case of negative helicity we can observe a peak in the density of the states and a shift of the latter towards the left as the parameter $\alpha_{SO}$ increases. This confirms what has been commented for the figures (\ref {fig2}, \ref {fig3}, \ref {fig4}, \ref {fig5}), or that the effects of a Rashba spin-orbit coupling become more marked for the negative helicity subband.

In a generic 3D system with free-electron like dispersion relation, the DOS  behaves as  the square root of energy,  whereas in a quantum layer ($2D$) the DOS is constant and so that it  jumps sharply every time a new quantum number from a new layer  takes over. In present case, the DOS shows almost a $2D-like$ behavior for the first subband. In fact, this is nearly pure 2D subbands with a negligible transversal hopping between the layers. Instead, at the bottom of the second subband appears a sharp step due to contribution of the partial density of states of the second subband to the total density of states. The total energy dispersion of the second subband, $\Delta E_{z2}$ determines the energy separation between the top and the band edge energy for the second subband. In the energy range $\eta_{2edge}<\eta < \eta_{2top}$ (where $\eta_{2edge}$ and $\eta_{2top}$ are, respectively, the Lifshitz parameter at the edge and at the top of the second subband), the electronic structure is like that of an anisotropic 3D electron gas, while the 2D character appears at higher energy $\eta>\eta_{2top}$.

 The observed Rashba shift can be understood  by considering the simplified model previously introduced. In the absence of RSOC the vanishing of the energy gradient occurs in an isolated point and, therefore, there are no singularities in the DOS. In contrast in the presence of RSOC, the energy gradient vanishes at finite values of the absolute values in-plane momentum, $ k_\parallel $, therefore we have singular points distributed on circumferences that generate van Hove peaks in the DOS.  As the RSOC increases, the energy at which van Hove peaks occur moves to the left, but, as underlined in the discussion of the tight-binding model, the difference in energy between the band edge and the maximum DOS value remains constant and equal to the dispersion along z.

The shape of the Fermi surface is crucial for understanding the electronic properties of metals. As first noticed by Lifshitz [\onlinecite{lifshitz1960anomalies}], changes in the Fermi surface topology cause anomalous behavior of thermodynamic, transport and elastic
properties of materials. Intuitively, the simplest way to observe such an electronic topological transition, also known as Lifshitz transition, is by tuning the
Fermi level to the singular point in the band structure where the change of topology takes place. This requires considerable variations of the electron density. A quantum critical point appears in the proximity of a Lifshitz transition with  typical quantum criticalities and possible quantum tricritical behavior in itinerant electron systems. 

There are two types of Lifschitz transition: type I, the appearing of a new detached Fermi surface region or appearing or disappearing of a new Fermi surface (FS) spot, and type II, the disrupting or the neck-collapsing-type of Lifschitz transition with a change of dimensionality that can be induced by orbital symmetry breaking in lightly hole doped bands. In the Fig.(\ref{fig4}) we show that a new 3D FS opens when the chemical potential crosses the band edge energy, and the electron gas in the metallic phase undergoes an electronic topological transition (ETT). When the chemical potential is beyond the band edge in an anisotropic system at a higher energy threshold, the electronic structure undergoes a second ETT, the 3D-2D ETT, where the FS changes topology from 3D to 2D or vice versa, called also the opening or closing of a neck in a tubular FS or neck collapsing.
This ETT is a common feature of all existing high-temperature superconductors and novel materials synthesized by material design in the search for room temperature superconductivity. 

However, the analysis made in this section highlights some surprising results, the first is that there is a change in symmetry of the evolution of the FS topology in passing from the first to the second subband. The second is that in the proximity of a  type II Lifshitz transition we have a curve of critical points and no longer an isolated point, a fact   which explains the appearance of a very pronounced peak in DOS values. The radius of this curve increases with the intensity of the RSOC and this is reflected in an increase and at the same time a true right shift of the DOS maximum. In this situation, the variation in the Fermi surface (FS) topology is absolutely non-trivial. 

 Clearly, we want to see how the above features of the electron spectrum and DOS are reflected in the properties of the superconducting phase. Therefore, after having analysed in detail the structure of the FS and the DOS in the normal phase, we turn, in the next section, to the study of the superconducting phase.

\section{The superconducting phase}

To investigate how the shape of the Fermi surface and the behavior of the DOS manifest in the superconducting properties of the system with RSOC, we first derive the equations for computing the energy gap. The approach used is the one originally  introduced by D. Innocenti \textit{et al}. [\onlinecite{innocenti2010resonant}] and successively developed in Refs.[\onlinecite{shanenko2012atypical}]-[\onlinecite{innocenti2010shape}],  where the Bogoliubov equations are solved analytically and numerically without the typical approximations of the BCS theory. The entirely new thing in the following discussion, however, consists in using non-relativistic Dirac wave functions in order to take into account the additional spin degree of freedom.

The field operators of Eq.(\ref{eq:interaction}) can be written in terms of the single-particles states
\begin{equation}
\psi_{n\mathbf{k}\alpha}\left(\mathbf{r}\right)=\varphi_{nk_{z}}\left(z\right)\frac{e^{i\mathbf{k}_{\parallel}\cdot\mathbf{r}_{\parallel}}}{\sqrt{\mathcal{A}}}\boldsymbol{\chi}_{\alpha}\equiv \tilde{\psi}_{n\mathbf{k}}\left(\mathbf{r}\right)\boldsymbol{\chi}_{\alpha},\label{eq:wavefunctionsforfieldop}
\end{equation}

where $\boldsymbol{\chi}_{\alpha}$ with $\alpha=\uparrow,\downarrow$ are the usual spinors associated to the quantization of the spin along the z axis. This is a legitimate expression for the field operators since the functions  $\psi_{n\mathbf{k}\alpha}\left(\mathbf{r}\right)$ are the eigenfunctions of the Hamiltonian $H_0$ obtained by setting $\alpha=0$ in Eq.(\ref{eq:h0}), i.e. completely neglecting the Rashba term. If we indicate with $c_{n\mathbf{k}\alpha}$ the operators that destroy a particle in the state (\ref{eq:wavefunctionsforfieldop}) then the field operators become

\begin{equation}
\Psi_{\alpha}\left(\mathbf{r}\right)=\sum_{n,\mathbf{k}}\psi_{n\mathbf{k}\alpha}\left(\mathbf{r}\right)c_{n\mathbf{k}\alpha}
\end{equation}

and the interaction term can be written as: 

\begin{equation}
\begin{split}
H_{I} =& \frac{U_0}{2}
\sum_{\mathbf{k}_{1},\mathbf{k}_{2},\mathbf{k_{3}},\mathbf{k_{4}}}
\sum_{\alpha,\beta} 
	I^{n_1,\mathbf{k}_{1};n_2,\mathbf{k}_{2}}_{n_3,\mathbf{k}_{3};n_4,\mathbf{k}_{4}} \ \ \cdot\\
&\cdot\ \ 	c^\dagger_{n_1,\mathbf{k}_{1},\alpha}
	c^\dagger_{n_2,\mathbf{k}_{2},\beta }
	c        _{n_3,\mathbf{k}_{3},\beta }
	c        _{n_4,\mathbf{k}_{4},\alpha}
\end{split},
\end{equation}
with the overlap integrals defined by
\begin{equation}
\begin{split}
	I^{n_1,\mathbf{k}_{1};n_2,\mathbf{k}_{2}}_{n_3,\mathbf{k}_{3};n_4,\mathbf{k}_{4}} =&
\int
	\tilde{\psi}^\ast_{n_1,\mathbf{k}_{1}}\left(\mathbf{r}\right)
	\tilde{\psi}^\ast_{n_2,\mathbf{k}_{2}}\left(\mathbf{r}\right)\ \cdot \\
& \ \ \cdot	\tilde{\psi}_{n_3,\mathbf{k}_{3}}\left(\mathbf{r}\right)
\tilde{\psi}_{n_4,\mathbf{k}_{4}}\left(\mathbf{r}\right)
	d\mathbf{r}.
	\end{split}
\label{eq:exchange1}
\end{equation}

The integrals in Eq.(\ref{eq:exchange1}) appear in the treatment of the superconductive phase transition in the presence of a periodic potential and have been extensively discussed [\onlinecite{innocenti2010resonant}].
The operators
$a^\dagger_{n,\mathbf{k},\lambda}$
that create a particle in the state Eq.(\ref{eq:wavefunction}) are related to the 
$c^\dagger_{n,\mathbf{k},\alpha}$
operators by an unitary transformation
\begin{equation}
a^\dagger_{n,\mathbf{k},\lambda} =
\sum_{\alpha} {
	c^\dagger_{n,\mathbf{k},\alpha}
	U_{\alpha,\lambda}\left(\mathbf{k}\right),
}\label{unitary_transformation}
\end{equation}
where the matrix element of the change of basis  is equal to
$
U_{\alpha,\lambda}\left(\mathbf{k}\right) =
\boldsymbol{\chi}^\dagger_\alpha \cdot
\boldsymbol{\eta}_{\lambda}\left(\theta_\mathbf{k_\parallel}\right).
$
As a result, the four operator products that appear in the expansion of the right hand side of Eq.(\ref{eq:interaction}) can be written as

\begin{eqnarray}
\begin{split}
& c^\dagger_{n_1,\mathbf{k}_{1},\alpha}
c^\dagger_{n_2,\mathbf{k}_{2},\beta }
c        _{n_3,\mathbf{k}_{3},\beta }
c        _{n_4,\mathbf{k}_{4},\alpha} 
=\\ 
& \sum_{\lambda_1,\lambda_2,\lambda_3,\lambda_4} 
	M_{\lambda_1,\lambda_4}\left(\theta_\mathbf{k_{1\parallel}} - \theta_\mathbf{k_{4\parallel}}\right)
	M_{\lambda_2,\lambda_3}\left(\theta_\mathbf{k_{2\parallel}} - \theta_\mathbf{k_{3\parallel}}\right)\ \ \cdot \\
& \ \ \cdot	a^\dagger_{n_1,\mathbf{k_{1}},\lambda_1}
	a^\dagger_{n_2,\mathbf{k_{2}},\lambda_2}
	a   _{n_3,\mathbf{k_{3}},\lambda_3}
	a          _{n_4,\mathbf{k_{4}},\lambda_4},
\end{split}
\end{eqnarray}
where we have defined 
\begin{equation}
M_{\lambda_1,\lambda_4 } (\theta_{ \mathbf{k}_{1\parallel}}-\theta_{\mathbf{ k}_{4\parallel}})=\sum_{\alpha} U^\dagger_{\lambda_1,\alpha}(\mathbf{k}_1)U_{\lambda_4,\alpha}(\textbf{k}_4),
\label{eq:m}
\end{equation}
and similarly for $M_{\lambda_2,\lambda_3}\left(\theta_\mathbf{k_{2,\parallel}} - \theta_\mathbf{k_{3,\parallel}}\right)$.
Since the $\varphi_{n,k_z}$ are Bloch wavefunctions, the integral (\ref{eq:exchange1}) is different from zero only for $\mathbf{k}_1+\mathbf{k}_2=\mathbf{k}_3+\mathbf{k}_4$ and the  expression for $H_I$ becomes:
\begin{equation}
\begin{split}
H_I=& \frac{1}{2}\sum_{n_1,n_2,n_3,n_4,\textbf{k}_1,\textbf{k}_2,\textbf{K}} U^{n_1,\lambda_1;n_2,\lambda_2}_{n_3,\lambda_3;n_4,\lambda_4}(\textbf{k}_1,\textbf{k}_2;\textbf{K})\\
&a^\dagger_{n_1,\textbf{k}_1,\lambda_1}a^\dagger_{n_2,-\textbf{k}_1+\textbf{K},\lambda_2}a_{n_3,-\textbf{k}_2+\textbf{K},\lambda_3}a_{n_4,\textbf{k}_2,\lambda_4},
\label{eq:interactionRSOCKP}
\end{split}
\end{equation}
where the effective potential reads
\begin{equation}
\begin{split}
&U^{n_1,\lambda_1;n_2,\lambda_2}_{n_3,\lambda_3;n_4,\lambda_4}(\textbf{k}_1,\textbf{k}_2;\textbf{K})=U_0 I^{n_1,\mathbf{k}_{1};n_2,-\mathbf{k}_{1}+\mathbf{K}}_{n_3,-\mathbf{k}_{2}+\mathbf{K};n_4,\mathbf{k}_{2}}\\
&M_{\lambda_1,\lambda_4}(\theta_{\mathbf{k}_{1\parallel}}-\theta_{\mathbf{k}_{2\parallel}})M_{\lambda_2,\lambda_3}(\theta_{-\mathbf{k}_{1\parallel}+\mathbf{K}_\parallel}-\theta_{-\mathbf{k}_{2\parallel}+\mathbf{K}_\parallel}).
\end{split}
\end{equation}

Equation (\ref{eq:interactionRSOCKP}) is the expression of the interaction term when both RSOC and a periodic potential are present. It can be viewed as the \textit{natural} extension to a multiband system of Eq.(4) of [\onlinecite{gor2001superconducting}] and, from this point on, the computation of the superconducting gap follows the same steps.
In agreement with Gor'kov and Rashba [\onlinecite{gor2001superconducting}] we assume that the normal and the anomalous Green functions are diagonal in the helicity base.
Hence we consider only Cooper pairs with zero net momentum ($\textbf{K}=0$), formed with particles in the same band and with the same helicity, $\{(n,\textbf{k},\lambda),(n,-\textbf{k},\lambda)\}$, which are connected by the time reversal symmetry operator. We allow for the contact exchange interaction $H_I$ to connect pairs in different bands with different helicity: the pair $\{(n,\textbf{k},\lambda),(n,-\textbf{k},\lambda)\}$ can be scattered into the pair $\{(l,\textbf{q},\nu),(l,-\textbf{q},\nu)\}$ where $n$, $\lambda$, $l$, and $\nu$ can assume any allowed value.  
We emphasize that, as discussed in  Ref.  [\onlinecite{gor2001superconducting}],   the existence of a different pairing function in each helicity band implies a mixture of singlet and triplet pairing. Symmetric and antisymmetric combinations (see Eq.(22) of  Ref.  [\onlinecite{gor2001superconducting}]) of the  pairing functions for the two helicity bands correspond to the singlet and triplet component with respect to the original spin quantization axis taken along the z direction. This can be seen by using the transformation (\ref{unitary_transformation}) connecting the electron operators between the original spin basis and the helicity basis.

Following the {\it standard} Gor'kov approach at finite temperature, we introduce the Matsubara imaginary time operators $a_{n,\textbf{k},\lambda} (\tau)$ that follow the imaginary time evolution equation $-{\partial_\tau} a_{n,\textbf{k},\lambda} (\tau)= [a_{n,\textbf{k},\lambda} (\tau), H ]$. In terms of these operators the {\it normal} $[G_{n,\lambda} (\textbf{k},\tau-\tau ' )]$ and the {\it anomalous} $[ F_{n,\lambda}^\dagger (\textbf{k},\tau-\tau ' )$ and $F_{n,\lambda}^\dagger (\textbf{k},\tau-\tau ' )]$ Green's functions are defined as
\begin{eqnarray}
G_{n,\lambda} (\textbf{k},\tau-\tau ' )&\equiv&-\langle T_\tau a_{n,\textbf{k},\lambda} (\tau) a^\dagger_{n,\textbf{k},\lambda} (\tau ')\rangle,\\
F^\dagger_{n,\lambda} (\textbf{k},\tau-\tau ' )&\equiv& \langle T_\tau a^\dagger_{n,-\textbf{k},\lambda} (\tau) a^\dagger_{n,\textbf{k},\lambda} (\tau ')\rangle,\\
F_{n,\lambda} (\textbf{k},\tau-\tau ' )&\equiv& \langle T_\tau a_{n,-\textbf{k},\lambda} (\tau) a_{n,\textbf{k},\lambda} (\tau ')\rangle,
\end{eqnarray}
where $T_\tau$ denotes the imaginary-time ordering operator.
By using a mean-field approach, we arrive, after a lenghty algebra, to the self-consistent gap equation:
\begin{equation}
\Delta_{n, \lambda}(\textbf{k}) =-\frac{1}{2} \sum_{l,\textbf{q},\nu} U'_{n,\lambda;l,\nu}(\textbf{k},\textbf{q})\frac{\Delta_{l,\nu}(\textbf{q})}{2E_{l,\nu}(\textbf{q})} tanh \bigg(\frac{\beta E_{l,\nu}(\textbf{q})}{2}\bigg),
\label{eq:selfcon1}
\end{equation}

	where $\Delta_{n, \lambda}(\textbf{k})$ is defined as

	\begin{equation}
	\Delta_{n, \lambda}(\textbf{k}) \equiv
		\frac{1}{2}
		\sum_{l,\textbf{q},\nu}
			U'_{n,\lambda;l,\nu}(\textbf{k},\textbf{q})
			F_{l,\nu}\left(\textbf{q},0^+\right),
	\label{eq:gapdefinition}
	\end{equation}

and the quasiparticle energy is
\begin{equation}
E_{l,\nu}(\textbf{q}) =\sqrt{(\varepsilon_{\nu, \mathbf{q}_{\parallel}}+\varepsilon_{l,q_z}-\mu)^2+|\Delta_{l,\nu}(\textbf{q})|^2}
\label{eq:12}
\end{equation}
and the pairing potential reads
\begin{equation}
\begin{split}
U'_{n,\lambda;l,\nu}(\textbf{k},\textbf{q}) &\equiv U^{n,\lambda;n,\lambda}_{l,\nu;l,\nu} (\textbf{k},\textbf{q};\textbf{0})-U^{n,\lambda;n,\lambda}_{l,\nu;l,\nu} (-\textbf{k},\textbf{q};\textbf{0})\\
&= U_0 I_{n,l}(k_z,q_z) \lambda \nu e^{-i(\theta_{\textbf{k}_\parallel}-\theta_{\textbf{q}_\parallel})},
\end{split}
	\label{eq:u}
\end{equation}
with the overlap integral
\begin{equation}
\label{overlap}
I_{n,l}(k_z,q_z)\equiv I^{n,\mathbf{k};n,-\mathbf{k}}_{l,\mathbf{q};l,-\mathbf{q}}.
\end{equation}

\begin{figure}
	\includegraphics[scale=0.43]{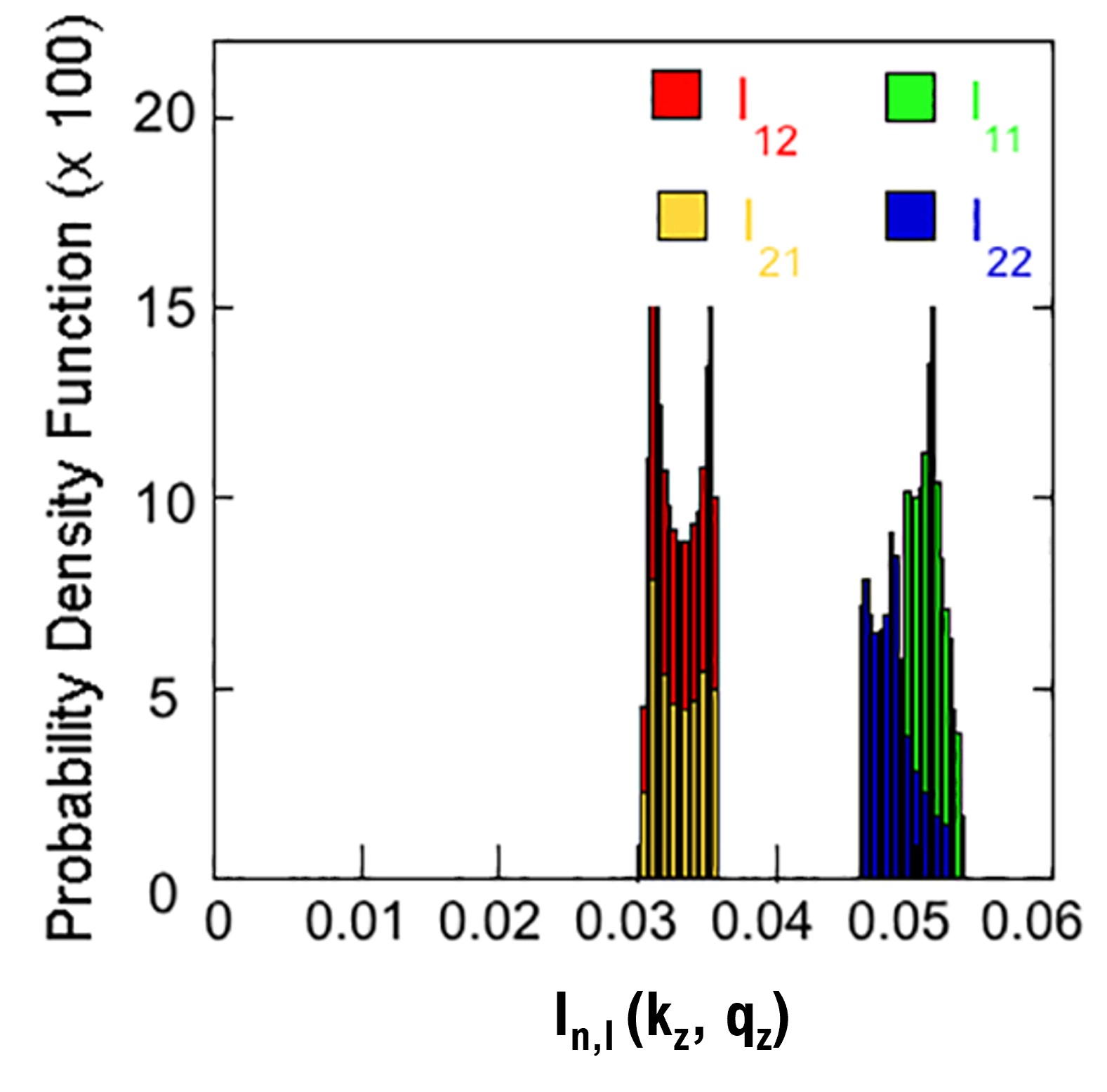}
	\caption{\foreignlanguage{english}{\textit{Histogram of the matrix elements defining the superposition integral of} Eq.(\ref{overlap}). The green and blue bars refer, respectively, to the intraband pairings $ I_ {1,1} (k_z, q_z) $ and $ I_ {2,2} (k_z, q_z) $, while the red and yellow bars refer to, respectively, to the interband couplings $ I_ {1,2} (k_z, q_z) $ and $ I_ {2,1} (k_z, q_z) $. The histogram shows a marked anisotropy.}\label{fig8}}
\end{figure}

The matrix elements defined in the Eq.(\ref{overlap})  depend on the subband index ($n$ and $l$) and on the wavevector transversal to the layers ($k_z$ and $q_z$). In the superposition integrals (Eq.(\ref{overlap})) only the dependence on the transverse moment remains, since the wave functions in the plane are plane waves which compensate for the choice made on $\mathbf{K}$. For a periodic potential barrier associated with the superlattice of layers the density histogram of pairing interaction matrix elements between subbands is illustrated in the Fig.(\ref{fig8}). The intraband (diagonal elements of matrix) and interband (off-diagonal elements of matrix) distributions show different shapes and widths and have different range of values. In particular, the off-diagonal elements have a probability density function which is about half of the diagonal elements which instead are of the same order of magnitude.

\begin{figure}
	\includegraphics[scale=0.3]{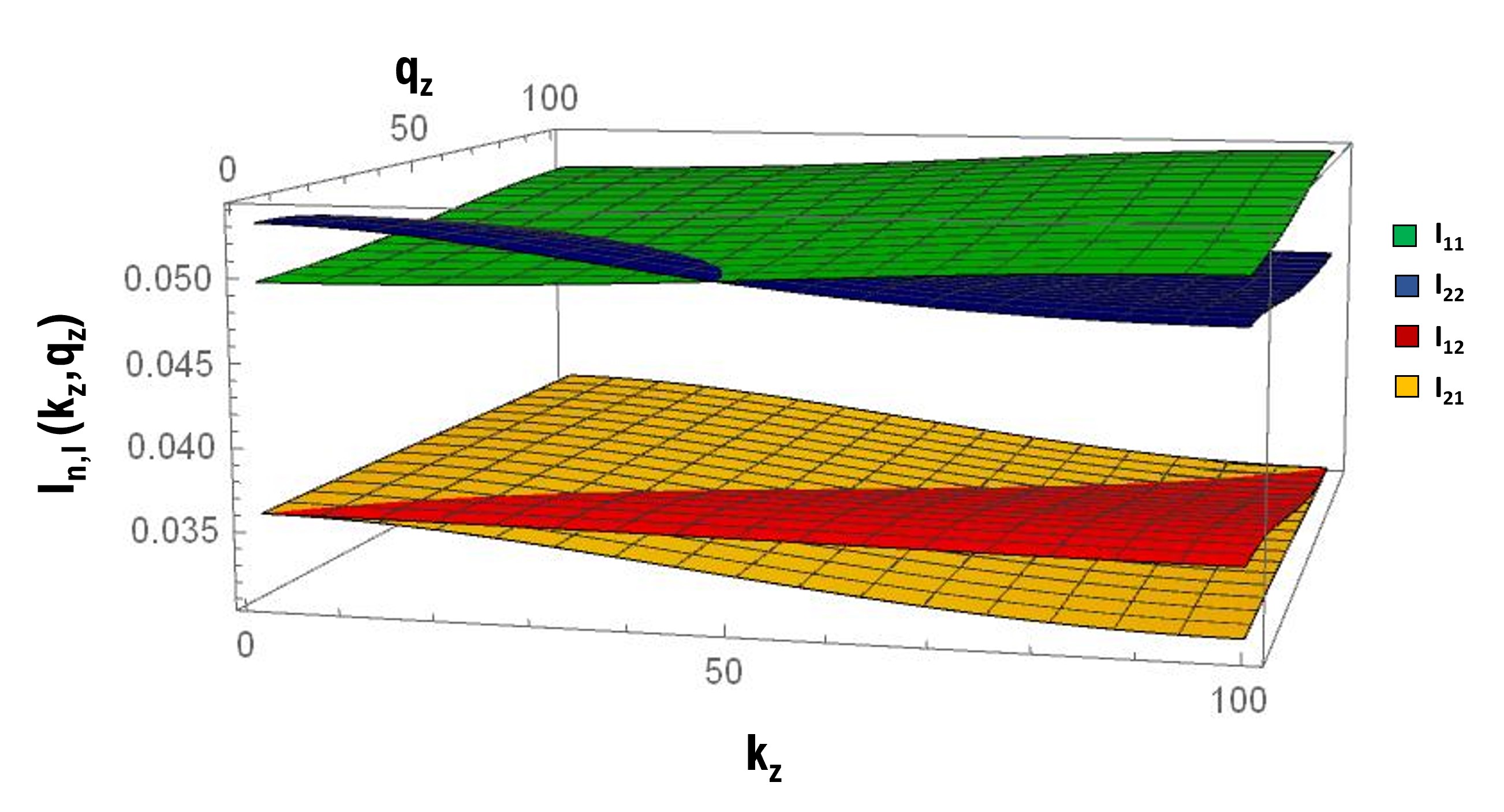}
	\caption{\foreignlanguage{english}{Terms of the matrix of the exchange integral defined in Eq.(\ref{overlap}) as a function of the wave vectors in the direction of the confinement potential. The green plan corresponds to $ I_ {1,1} (k_z, q_z) $, the blue plan corresponds to $ I_ {2,2} (k_z, q_z) $, the red plan corresponds to $ I_ {1,2} (k_z, q_z) $ and the yellow plan corresponds to $ I_ {2,1} (k_z, q_z)$.}\label{fig9}}
\end{figure}

While in the Fig.(\ref{fig8}) the dependence on the band indices of the exchange integral is highlighted, in the Fig.(\ref{fig9}) the dependence on wave vectors is highlighted. This last figure clearly shows that the diagonal elements of the matrix defined by the superposition integral are greater than those off-diagonal, whatever the value of the wave vectors. Furthermore, for both the intraband and the interband there is a curve of values of $ k_z $ and $ q_z $ for which $ I_ {11} = I_ {22} $ and $ I_ {12} = I_ {21} $, whereas on the right of this curve $ I_ {11} <I_ {22} $ and $ I_ {12} <I_ {21} $,  the opposite being true on the left. 
 
The integral equation (\ref{eq:selfcon1}) shows a  dependence of the gap $\Delta_{n,\lambda}(\textbf{k})$, reminiscent of the Rashba spinors Eq.(\ref{eq:spinors}),  upon the helicity and the in-plane component of the wave vector through a phase factor $\lambda e^{i\theta_{\mathbf{k}_\parallel}}$.  To get rid of this dependence in the self-consistent equation, we define an auxiliary gap function $\Delta_n (k_z)$ as
\begin{equation}
\lambda e^{i\theta_{\mathbf{k}_\parallel}} \Delta_n(k_z) \equiv \Delta_{n,\lambda}(\textbf{k}).
\label{gapform}
\end{equation}

Then, the self-consistent equation for $\Delta_n(k_z)$ can be cast in the form

	\begin{equation}
	\begin{split}
	\Delta_{n}(k_z) &=-\frac{U_0}{2} \sum_{l,q_z} I_{n,l}(k_z,q_z) \Delta_l (q_z) \\
	&\times\sum_\nu \sum_{q_x,q_y} 
	\frac{
		\tanh \bigg(\frac{\beta}{2} E_{l,\nu}(\textbf{q})\bigg)
	}{
		{2E_{l,\nu}(\textbf{q})}
	},
	\label{eq:selfconsistency2}
	\end{split}
	\end{equation}

where it is understood that the wave vector $\mathbf{q}$ appearing in the last term is $\textbf{q}=(q_x,q_y,q_z)$.
The solution of Eq.(\ref{eq:selfconsistency2})
 is obtained numerically by starting with a guess for $\Delta_{n}(k_z)$ and iterating until convergence is reached. Since the computational effort is considerable, it becomes important to speed-up the calculations by reducing the dimensionality of the summations. In fact, once $q_z$ has been fixed, the argument of the last sum depends on $q_\parallel$ only trough  the in-plane dispersion energy $\varepsilon_{\nu,q_\parallel}$. It is, then,  convenient to define a \textit{partial} density of states $g_{\nu}(\varepsilon_\parallel)$ that allows a transformation of the double sum in Eq.(\ref{eq:selfconsistency2}) into a one-dimensional integral:

	\begin{equation}
	\sum_{q_x,q_y} f(\varepsilon_{\nu,q_\parallel})=\frac{{\cal A}}{4\pi^2} 	\int^{\varepsilon_{\parallel,max}}_{\varepsilon_{\parallel,min}}  g_\nu(\varepsilon_\parallel) f(\varepsilon_\parallel) d\varepsilon_\parallel.
	\end{equation}

 The integration extrema, $\varepsilon_{\parallel,min}$ and $\varepsilon_{\parallel,max}$, are computed by introducing the contact interaction energy cut-off in the sense that the condition $\varepsilon_{\parallel,min}<\varepsilon_{\parallel}<\varepsilon_{\parallel,max}$ implies the inequality $| \varepsilon_{\parallel}+\varepsilon_{l,q_z}-\mu |<\hslash\omega_0$.

	It is worth to point out that $g_\nu(\varepsilon_\parallel)$ cannot be formulated as a
	single analytical function but it has to be defined with a piecewise expression that
	reflects the topology change of the Fermi Surface when switching from one regime to
	another (see sections III.A, III.B, and III.C).
	In fact, the expression defining the partial density of states is:

		\begin{equation}
		\begin{split}
			\frac{\mathcal{A}}{4\pi^2} g_\nu(\varepsilon_\parallel) &= 
			\sum_{q_x,q_y} \delta(\varepsilon_{\parallel} - \varepsilon_{\nu,\mathbf{q}_\parallel}) \\
			&= \frac{\mathcal{A}}{4\pi^2} \int^{\infty}_{0} 
				2 \pi \mathbf{q}_{\parallel} \delta\left( \varepsilon_\parallel - \varepsilon_{\nu,q_\parallel} \right) d\mathbf{q}_\parallel
		\end{split}
		\end{equation}

	where the double sum has been transformed in a integral in polar coordinates in the last line.
	This leads to the following expression for $g_\nu(\varepsilon_\parallel)$:
		\begin{equation}
			g_\nu(\varepsilon_\parallel) =
			\begin{cases}
				4 \pi m \frac{-2 \nu k_0 }{\sqrt{2 m \varepsilon_\parallel + k_{0}^2} } 
					&\text{if } -\frac{k^2_0}{2 m} \leq \varepsilon_\parallel < 0, \ \nu = -1 \\
				4 \pi m \frac{ \nu k_0 + \sqrt{2 m \varepsilon_\parallel + k_{0}^2} }{\sqrt{2 m \varepsilon_\parallel + k_{0}^2} } 
					&\text{if } \varepsilon_\parallel > 0 \\
				0 
					&\text{otherwise} \\
			\end{cases}
					\end{equation}
	where $k_0$ is defined as in the discussion preceding Eq.(\ref{eq:Brancab}), but this time we do not set $2m = 1$.

The Eq.(\ref{eq:selfconsistency2}) has been solved both in the limit $T\rightarrow 0$, (that is $\beta \rightarrow \infty$), and in the limit $T\rightarrow T_C$, (that is $\Delta_{n}(k_z)\rightarrow 0$  for every $n$ and $k_z$). The first limit allows to determine the gaps while the second allows to determine the critical temperature.

The results of the numerical computations for the gaps are shown in the Fig.(\ref{fig10})A, in which we plot both partial DOS and $\Delta_n(k_z)$ for the first and second subbands in $ k_z = \pi / 2d $ as a function of the Lifshitz parameter rescaled, $\eta_R$ (Eq.(\ref{eq:lifr})). The numerical values of the shift due to the Rashba coupling are indicated in the various panels which differ in the value of the $ \alpha_ {SO} $ parameter.
Furthermore, in this discussion, we set the value of the superconducting coupling at $ g = 0.4 $, where $ g $ is defined as $ g = g_ {3D} (\mu) U_0$ with $ g_ {3D} = \frac{1}{(2\pi)^2(\sqrt{\hslash^2/2m})^3 } \sqrt{\mu}$ being the DOS at the Fermi level for a homogeneous system (no RSOC, no peridodic potential along $z$). In the numerical simulation we assume that $ g $ is a constant, so as the chemical potential changes both $ g_ {3D} $ and $ U_0 $ are continuously recalculated.

\begin{widetext}
	
\selectlanguage{american}%
\begin{figure}
	\includegraphics[scale=0.5]{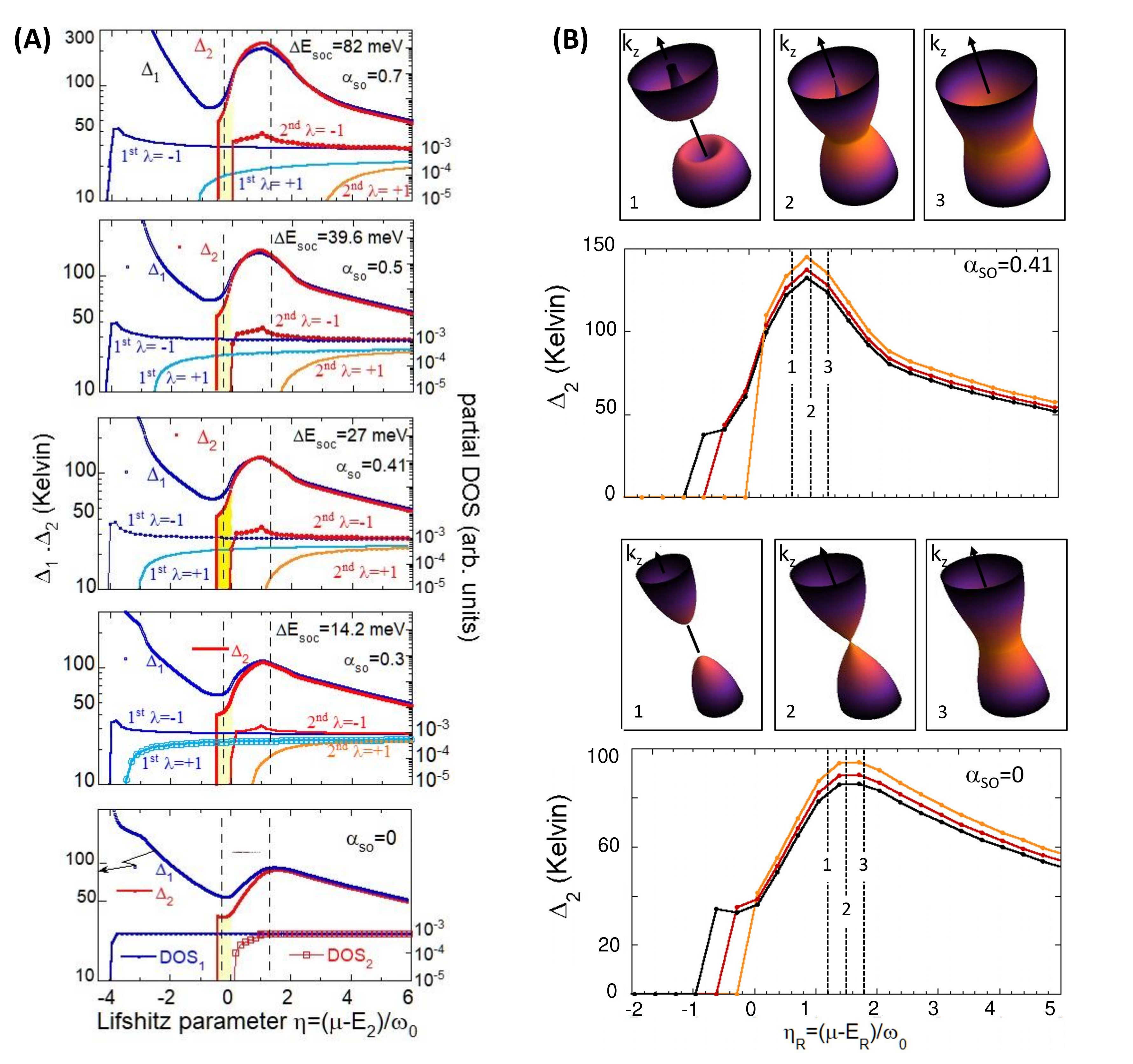}
	\caption{\foreignlanguage{english}{\textit{Properties of the normal phase and the superconductive phase vs the  rescaled Lifshitz parameter without and with RSOC for different $\alpha_{SO}$ values such that $\Delta E_{RSOC}\gtreqqless \Delta E_{2z}$}.
		\textit{Panel A} left side: starting from the bottom, the first panel shows the DOS for the first subband (blue curve) and the second subband (red curve) and the trend of the gap relative to the first subband (blue curve) and to the second subband (red curve) as a function of the  rescaled Lifshitz parameter. 
		The other panels in (A) are related to the four values of $\alpha_{SO}$ previously discussed, $\alpha_{SO}=0.30,\ 0.41,\ 0.50,\ 0.70$, and show the partial DOS for the first subband with positive helicity (light blue curves) and with negative helicity (blue curves) and the partial DOS for the second subband with positive helicity (orange curve) and with negative helicity (red curves). We also report the trend of the gap relative to the first subband (blue curve) and to the second subband (red curve) as a function of the  rescaled Lifshitz parameter. 	
		An anomalous behavior and an amplification of the parameters of the superconductive phase are observed in a range of rescaled Lifshitz parameter $ 0<\eta_R < 1 $. i.e., in the proximity to the unusual van Hove singularity.
	\textit{Panel B} right side: the values of the gaps for the second subbands at $ \alpha_ {SO} =0 $ and $ \alpha_ {SO} =0.41 $ \textit{versus} the rescaled Lifshitz parameter for different values of $ k_z = 0, \ \pi / 2d, \ \pi / d$ show a small variation in a neighborhood of van Hove unusual singularity. We have highlighted the variation on the Fermi surface for three values of the Lifshitz parameter in $\eta_R = \eta_L $ and  $\eta_R = \eta_L \pm 0.5$ by choosing the black color for $ k_z =  \pi / d $, the red color for $ k_z = \pi / 2d $ and the orange color for $ k_z = 0 $ to show the trend of the charge of the gap values at different $k_z$. }}
	\label{fig10}
\end{figure}

\end{widetext}

 We emphasize that, in order to have the full gap, i.e. $\Delta_{\lambda n} ({\mathbf k})$, we must also consider the dependence on the phase factor and on the helicity, for this purpose we keep in mind the Fig.(\ref{fig5}).
 
 The Fig.(\ref{fig10})A shows that both for the gap of the first subband ($ \Delta_1 $) and for the gap of the second subband ($ \Delta_2 $) it is possible to distinguish three distinct regimes of multigap superconductivity as a function of the rescaled Lifshitz parameter when is tuned around the unusual van Hove singularity: an antiresonance regime in which the gaps reach a minimum value for $ \eta_R <\eta_L$, where $\eta_L $ is the value of the van Hove energy for which the DOS shows a peak, a resonance regime for $ \eta_R =\eta_L $ in which the gaps reach their maximum value and, finally,  a multi-band BCS-like regime for $ \eta_R >\eta_L $.

In particular, it can be observed that $ \Delta_1 $ has a minimum when the chemical potential is near the bottom of the second subband. The partial DOS relative to the first subband, both for $\lambda = 1$ and for $\lambda = -1$,  does not change as the chemical potential changes, therefore, the presence of such a pronounced minimum may be due to the existence of a Fano-type antiresonance in superconducting gaps. An antiresonance can be due to an interband exchange term that generates interference effects between the wave functions of a single particle by coupling in a non-trivial way the parameters of the superconducting phase relating to different bands. Both the depth and the position of the minimum in the $ \Delta_1 $ depend on this term.

A minimum in $ \Delta_1 $ appears below the band edge where the DOS of the second subband changes abruptly and the Fermi surfaces, as seen above, are in a Lifshitz transition of the first type. That is, the partial filling of the second subband is reflected in the appearance of two new three-dimensional (3D) Fermi surfaces, one for each helicity.

As for the gap of the second subband ($ \Delta_2 $), the Fig.(\ref{fig10})A shows that it starts to assume non-zero values when the chemical potential has not yet reached the bottom of the second subband. This effect emphasizes, once again, the non-banal role of interband coupling in a multicomponent system. 

$ \Delta_2 $ reaches the maximum corresponding to the maximum of the partial DOS relative to the second subband and to a negative helicity, i.e. when the chemical potential is near the unusual van Hove singularity, in which the Fermi surfaces changes topology passing from a 3D to a two-dimensional (2D) geometry. As the Rashba parameter $ \alpha_ {SO} $ varies, as seen previously, the radius of the circumference of the singular points that characterizes the Fermi surface in a Lifshitz transition of the type II (3D-2D ETT) increases, and, as shows the Fig.(\ref{fig10})A, the maximum values of $ \Delta_1 $ and $ \Delta_2 $ also increase. By varying the parameter $ \alpha_ {SO} $, we distinguish three different regimes: if $ \alpha_ {SO} $ is such that $ \Delta E_ {RSOC} <\Delta E_ {z2} $ the maximum of $  \Delta_1 $ has a value greater than the maximum of $ \Delta_2 $, for $ \Delta E_ {RSOC} = \Delta E_ {z2} =\omega_0 $ the maxima of the two gaps coincide within the limits of the numerical approximations made and, finally, for $ \Delta E_ {RSOC}> \Delta E_ {z2} $ the maximum of $ \Delta_2 $ exceeds the value of the maximum of $ \Delta_1 $.

It can also be noted that in the high energy limit the values of the gaps are to  a good approximation close to the BCS limit, i.e.  in the high energy limit the gaps no longer depend on $ \alpha $ [\onlinecite{gor2001superconducting}].

In Fig.(\ref{fig10})B, we plot the values of $ \Delta_2 $ as a function of $ \eta_R $ for different values of $ k_z $. It can be observed that $ \Delta_1 $ does not vary as $ k_z $ varies from point $ \Gamma $ to point $ Z $ of the IBZ, while it is possible to notice a small variation of $ \Delta_2 $ in a neighborhood of $ \eta_L $, where the role of exchange integrals (Eq.(\ref{eq:exchange1})) becomes crucial. 

For values of the Lifshitz parameter close to the van Hove singularity, for the second subband and for a helicity $ \lambda = -1 $ (the only one present) we plot  the corresponding FS highlighting the dependence of $ \Delta_2 $ from $ k_z $ with three different colors. In proximity of the unusual van Hove singularity the gap is not constant in $ k_z $ since the partial filling of the second subband causes the weight of $ \varepsilon_{l, q_z} $ in the Eq.(\ref{eq:12}) to be not negligible.

By solving the Eq.(\ref{eq:selfconsistency2}) in the limit $\Delta_{n}(k_z)\rightarrow 0$  we can be compute the critical temperature, $T_C$.

\begin{figure}
	\includegraphics[scale=0.5]{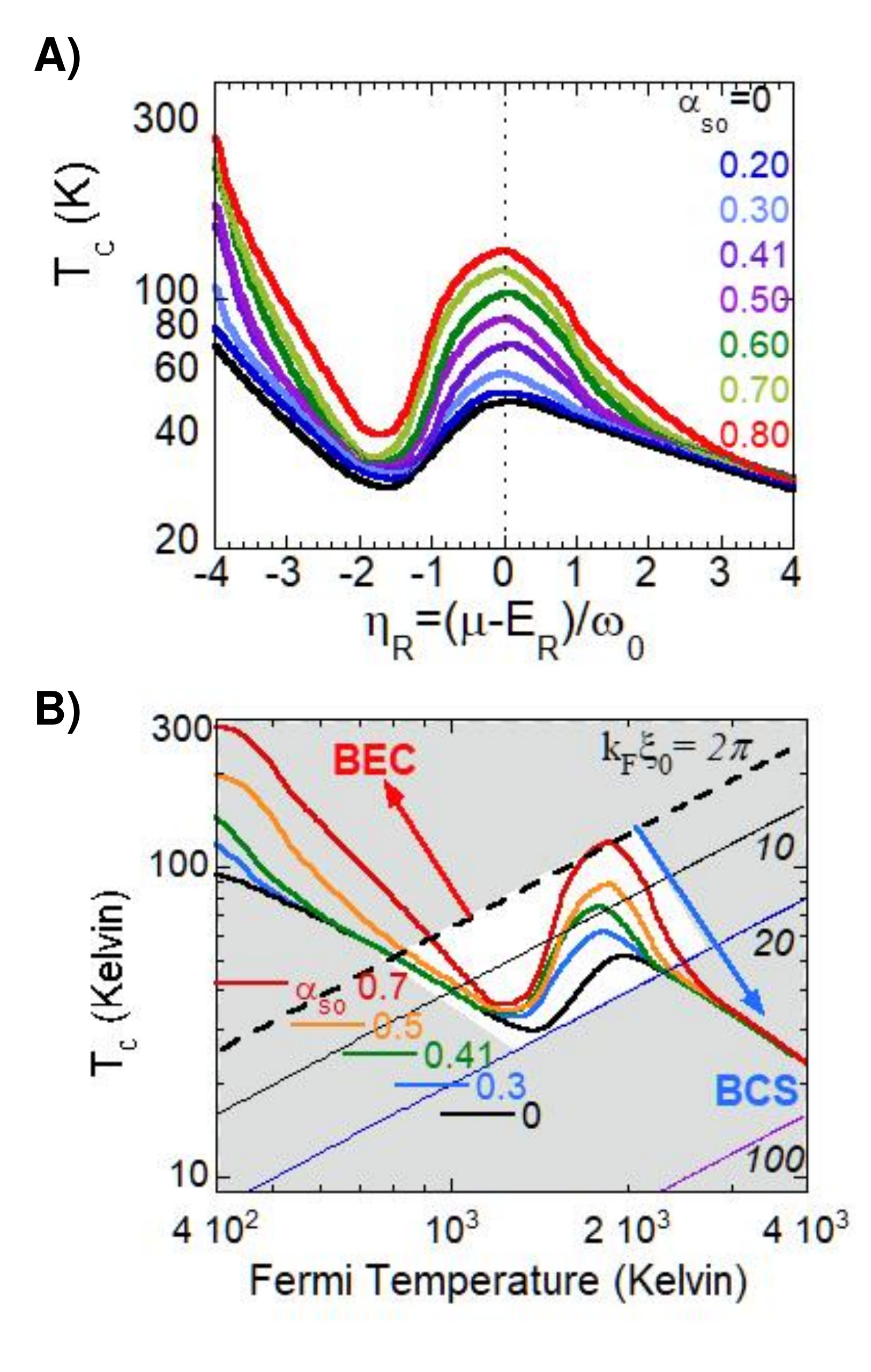}
	\caption{\foreignlanguage{english}{\textit{Panel A}: the critical temperature $T_C$ versus Lifshitz parameter for different values of Rashba coupling $\alpha_{SO}$ on a semilogarithmic scale. The critical temperature appears as an asymmetric function of the Lifshitz parameter and if $ \alpha_ {SO} <0.41 $ the maximum of $ T_C $ grows slowly, while if $ \alpha_ {SO}> 0.41 $ it grows faster and faster. 
\textit{Panel B}:  In this Uemura plot the critical temperature $T_C$ is plotted on a log-log scale versus the Fermi temperature for different values of Rashba coupling $\alpha_{SO}$. The white  box refers to the Fano resonance appearing near the BEC-BCS crossover indicated by the dashed line.}}\label{fig11}
\end{figure}

In the Fig.(\ref{fig11}), panel A, we plot the values of the critical temperature at different values of the Rashba parameter, $ \alpha_ {SO} $, as a function of the rescaled Lifshitz parameter. The critical temperature appears as an asymmetric function of the Lifshitz parameter regardless of the value of the $ \alpha_ {SO} $ parameter and shows the typical trend of a Fano antiresonance with a minimum at the first Lifshitz transition and a maximum at the second Lifshitz transition where the Fermi surfaces switch from 3D geometry to 2D geometry. From the Fig.(\ref{fig11})A  one can observe that in the presence of RSOC the energies are shifted to the left by an amount equal to $ E_0 =-( m \alpha^ 2)/(2\hbar^2)$ and that the values of the $T_C$ are amplified with respect to the case in which there is no RSOC. In particular, a maximum $T_C$ value is observed in correspondence with the van Hove singularity in the DOS because we have assumed  the energy cut off and the energy dispersion in the z direction to be the same.  The BCS theory predicts a value of about $32\ Kelvin$ for the critical temperature, with the model parameters chosen in this work, for $ \alpha_ {SO} = 0.4 $ this value increases about four times. 

In the \textit{panel} B of Fig.(\ref {fig11})  we show in a log-log plot the critical temperature $T_C$ as a function of the effective Fermi temperature 
$ T_F = E_F / k_B $ where the Fermi level is calculated from the bottom of the first subband and $ k_B $ is the Boltzmann constant. 
The critical temperature is calculated for different values of the Rashba coupling constant, $  \alpha_ {SO} = 0, \ 0.30, \ 0.41, \ 0.50, \ 0.70 $. The Fano resonance at the bottom of the second subband occurs in this so called Uemura plot [\onlinecite{uemura1997bose}] $T_C$  versus $T_F$.
 In this figure the dashed line indicates the BEC-BCS crossover predicted to be $ T_C = T_F /  (k_F  \xi_0)$  [\onlinecite{andrenacci1999density,perali2004quantitative,pistolesi1994evolution}].
  The Fano resonance clearly occurs on the BCS side of the BCS-BEC crossover where the ratio between $ T_F $ and $ T_C $  is in the range between 10 and 20.
   The calculated Fano resonance in the white region occours on the BCS side up to the largest spin-orbit coupling. In fact the Fano resonance occurs in the range  between the BEC crossover  and  the line  $ T_C = T_F / 20 $  in the BCS side. 
   From the figure it can be seen that the critical temperature values remain included in a BCS regime although as $ \alpha_ {SO} $ increases the Fano resonance appears increasingly shifted towards the BEC limit.

\selectlanguage{american}%

\begin{figure}
	\includegraphics[scale=0.6]{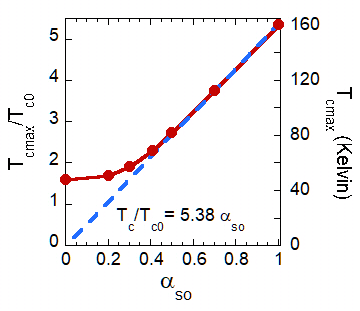}
	\caption{\foreignlanguage{english}{\textit{The maximum value of the critical temperature (red curve) and critical temperature predicted by the BCS theory ratio for different values of $ \alpha_ {SO}$ parameter}. What is observed is a marked amplification of The maximum of the $T_C$ when a Rashba coupling is introduced into the system. In particular, the maximum of the critical temperature increases slowly for $ \alpha_ {SO}<0.41$ and always becomes faster for $ \alpha_ {SO}>0.41$. }}\label{fig12}
\end{figure}

Furthermore, it is possible to observe that the value of $ \alpha_ {SO} $ for which $ \Delta E_ {RSOC} = \Delta E_ {z2} = \omega_0 $ i.e. $ \alpha_ {SO} = 0.41 $ marks the boundary between two distinct situations: if $ \alpha_ {SO} <0.41 $ the maximum of $ T_C $ grows slowly, while if $ \alpha_ {SO}> 0.41 $ it grows faster and faster. All this is highlighted in the Fig.(\ref {fig12}) in which we report the maximum of the $T_C$ as a function of the Rashba coupling constant (red curve). The maximum of critical temperature increases linearly with RSOC for $\alpha_{SO}\geq0.41$.

\selectlanguage{american}%

\begin{figure}
	\includegraphics[scale=0.8]{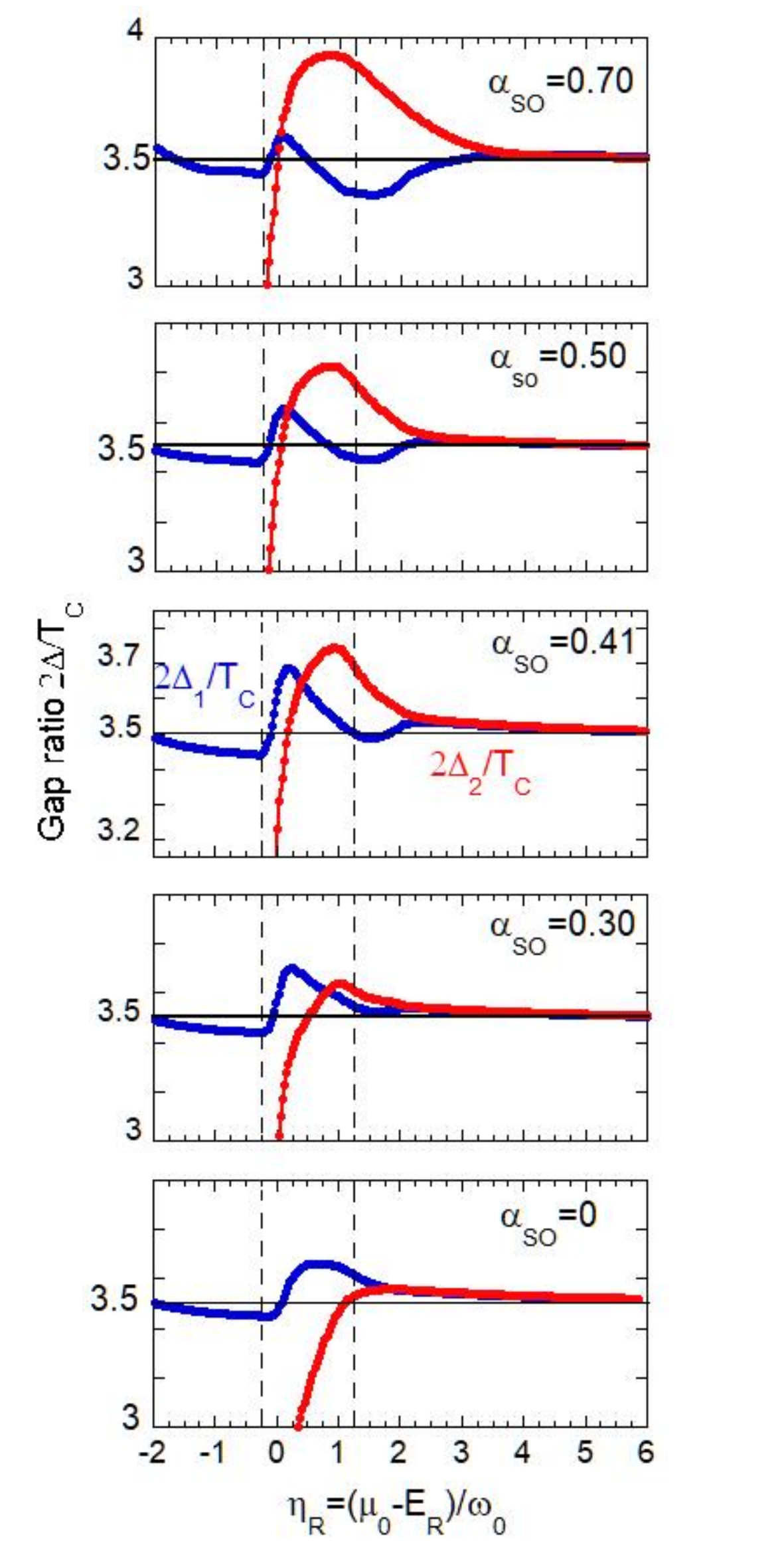}
	\caption{\foreignlanguage{english}{\textit{Properties of the superconductive phase vs the Lifshitz parameter without and with RSOC for different $\alpha_{SO}$ values such that $\Delta E_{RSOC}\gtreqqless \Delta E_{2z}$}. This figure shows both the trend of the gap ratio for the first subband (blue curves) and the second subband (red curves) compared to the constant value predicted by the BCS theory (black curves).}}\label{fig13}
\end{figure}

Previously we stressed the fact that near $ \eta_L $ the gaps vary with $ k_z $, this being strongly reflected in the calculation of the gap ratio, $ 2 \Delta / T_C $. Therefore, in order to plot this parameter correctly we consider $ \Delta $ averaged over $ k_z $. So, starting from the bottom of the Fig.(\ref{fig13}) we plot the gap ratio, $2\Delta /T_C$, where $T_C$ is the critical temperature, for the first and the second subband for different values of the $\alpha_{SO}$ parameter as a function of the  rescaled Lifshitz parameter.

\selectlanguage{american}%

We observe that the gap ratio differs from the constant value $3.5$ foreseen by the BCS theory when the  rescaled Lifshitz parameter is closed to $0<\eta_R<1$. In particular, the $ 2 \Delta_1 / T_C $ ratio for the first subband reaches a minimum, lower than the value predicted by BCS theory, when the rescaled Lifshitz parameter is approximately equal to zero. That is, when $ \Delta_1 $ is in an antiresonance regime and the system is close to a Lifshitz transition of the first type. The $ 2 \Delta_1 / T_C $ ratio  reaches a maximum value for $\eta_R \approx 1$, when the superconducting parameter $\Delta_1$ is in a resonance regime, this occurs close to the type II Lifshitz transition.

Regarding to the gap ratio for the second subband, $ 2 \Delta_2 / T_C $, we observe a significant deviation from the value predicted by the BCS theory in a range of values of the rescaled Lifshitz parameter equal to $ 0 <\eta <1 $. In particular, when the system is in an antiresonance regime $ 2 \Delta_2 / T_C $ diverges, while when $ \Delta_2 $ is in a resonance regime it shows a maximum. 
 By contrast, such a maximum is not present in the absence of the RSOC as the bottom panel of Fig. \ref{fig13} shows. 
As the parameter $ \alpha_ {SO} $ changes, the maximum of $ 2 \Delta_2 / T_C $ increases and, as in the case of the Fig.(\ref {fig10}), we observe three distinct regimes: when $ \Delta E_ {RSOC} <\Delta E_ {z2} $ we have $ 2 \Delta_2 / T_C <2 \Delta_1 / T_C $, when $ \Delta E_ {RSOC} <\Delta E_ {z2} = \omega_0 $ the two gap ratios intersect and, finally, for $\Delta E_ {RSOC}> \Delta E_ {z2} $ we have $ 2\Delta_2 / T_C> 2 \Delta_1 / T_C $.

\begin{figure}[h]
	\includegraphics[scale=0.6]{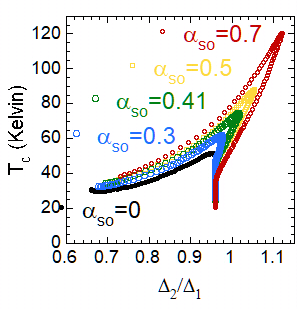}
	\caption{\foreignlanguage{english}{\textit{The trend of the critical temperature versus the $ \Delta_2 / \Delta_1 $ ratio for different values of the parameter $ \alpha_ {SO} $}. The $ T_C $ trend shows a strong asimmetry which becomes maximum when the $ \Delta_2 / \Delta_1 $  ratio is maximum.}}\label{fig14}
\end{figure}

 We see in Fig. \ref{fig13} that  the gap ratio to the transition temperature  $ 2 \Delta_2 / T_C <3.9$ in the second subband, at the maximum critical temperature, in spite of the peak of the partial DOS in the second subband due the van Hove singularity brought about by the  largest spin-orbit coupling  $\alpha_{SO}=0.7$, does not show a large deviation from the standard weak coupling universal value $3.52$ predicted by the single-band BCS theory. This is in agreement with the corresponding gap ratio $ 2 \Delta_1/ T_C = 3.4$ in the first subband. 
We plot $ T_C $ versus the $ \Delta_2 / \Delta_1 $ ratio for different values of the parameter $ \alpha_ {SO} $ in Fig. \ref{fig14}, which
 shows that  $ \Delta_2 / \Delta_1 $ ratio  is only  $1.12$, at maximum $T_C $,  for $\alpha_{SO}=0.7$.
Moreover we want to point out  that for  $\alpha_{SO}=0.41$ the gap ratio $ \Delta_2 / \Delta_1 <1$ while the ratio  between the partial DOS $  N_2 /  N_1 >1$ due to the van Hove singularity in the second subband.
These results show that the present superconducting scenario is in the weak coupling regime where the mean field approximation is valid.
In fact  the aim of this work is to show a scenario with weak electron-phonon coupling, where the amplification of the critical temperature has been driven by interband pairing in the presence of strong spin-orbit coupling. 
It is well known that in the multigap Bogoliubov superconductivity [\onlinecite{Bussmann2003,Bang2008,Dolgov2009}] the $ \Delta_2 / \Delta_1 $ ratio becomes proportional to  $  N_1 /   sN_2$  where the contact non retarded-exchange interaction (interband pairing) becomes more relevant that the retarded bosonic exchange pairing. From  Fig. \ref{fig14}  we can see a marked anisotropy in the trend of the critical temperature which shows a maximum corresponding to the maximum value of the $ \Delta_2 / \Delta_1 $ ratio.

Further work is in progress to study the cooperative role of contact  and retarded interactions in anisotropic superconductivity related with the anisotropic k-space pairing in the Fermi surface topology at unconventional Lifshitz transitions.
% relate the trend of the critical temperature with the values obtained for the gaps, we plot $ T_C $ versus the $ \Delta_2 / \Delta_1 $ ratio for different values of the parameter $ \alpha_ {SO} $, Fig.(\ref{fig14}). 

\selectlanguage{american}%

\begin{figure}
	\includegraphics[scale=0.7]{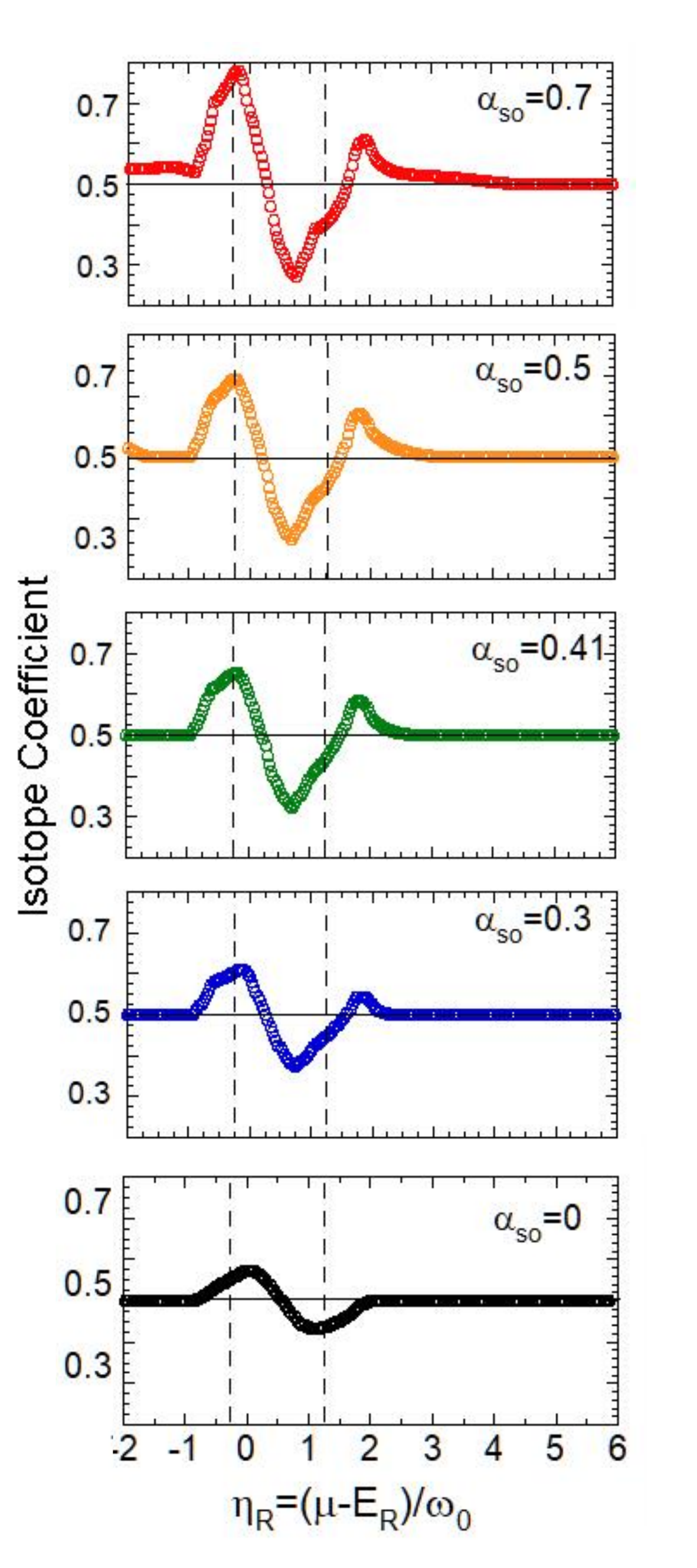}
	\caption{\foreignlanguage{english}{Properties of the superconductive phase vs the Lifshitz parameter without and with RSOC for different $\alpha_{SO}$ values such that $\Delta E_{RSOC}\gtreqqless \Delta E_{2z}$}. This figure shows the variation of the critical temperature with the cut-off energy via the isotope coefficinet. The constant value predicted by the BCS theory is the black line.}\label{fig15}
\end{figure}

As we have just seen, the  Fig.(\ref{fig10}) and the Fig.(\ref{fig13}) clearly show a quantum resonance characterized by a Fano-type asymmetry in the superconducting parameters and a considerable deviation from the predictions of the BCS theory. To further highlight this last aspect we graph the isotopic coefficient, $ \gamma = \partial ln T_C / \partial ln M $, as a function of the  rescaled Lifshitz parameter for different values of the parameter $ \alpha_ {SO} $, assuming that the cut-off energy depend on the isotopic mass as $\omega_0 \propto M^{-1/2}$ [\onlinecite{perali1997isotope}-\onlinecite{perali2012anomalous}] (Fig.(\ref{fig15})).

\selectlanguage{american}%

In the BCS theory, the isotope coefficient has a constant value as the chemical potential changes equal to $ 0.5 $, in our case instead we can notice a considerable deviation from this value when the rescaled Lifshitz parameter is in the range $ 0 <\eta_R <1 $ (for this range of values, the behavior of the $ \gamma $ parameter is that typical of the Fano antiresonance), that is, when the system is close to a Lifshitz transition. These deviations from the BCS theory increase as the Rashba coupling, $ \alpha_ {SO} $, increases, therefore there exists an unconventional dependence of the critical temperature on the cut-off energy unlike what is proposed in the BCS theory. 

In the high-energy limit, the gap ratio and the isotope coefficient tend to the values predicted by the BCS theory, so we are dealing with two BCS-like condensates.

\begin{figure}
	\includegraphics[scale=0.6]{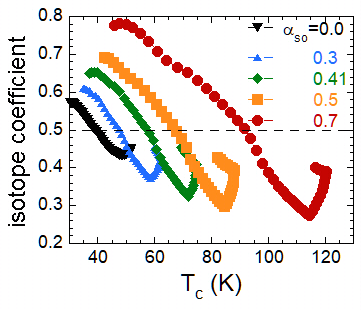}
	\caption{\foreignlanguage{english}{\textit{The isotope coefficient as a function of the critical temperature for different values of}} $ \alpha_ {SO} $.}\label{fig16}
\end{figure}

In Fig.(\ref{fig16}) we plot the isotope coefficient as a function of the critical temperature for different values of $ \alpha_{SO} $ for the range of energies delimited in Fig.(\ref{fig15}) by the dashed lines. This parameter, in this range of energies, can be measured and this prediction can be experimentally verified.

\selectlanguage{american}%

These results confirm that in correspondence with the van Hove singularity there is an amplification of the characteristic parameters of the superconductive phase which becomes more and more evident when the Rashba coupling exceeds a limit value of $ 0.4 $. 

The works [\onlinecite{innocenti2010resonant}-\onlinecite{innocenti2010shape}]  investigated the superconducting properties for a superlattice of quantum wells and observed that there is an optimum condition for the amplification of the critical temperature that is obtained when the cut-off energy is equal to the dispersion along the confinement direction of the higher energy band. The particular geometry considered creates, in fact, a multicomponent system. Here, instead, by introducing the degree of freedom of spin in the solution of the Bogoliubov equations, as well as having the possibility of dealing with realistic cases, we can overcome the limit imposed by previous works simply by suitably  increasing the Rashba coupling that exists by definition at the interface between different materials that make up a heterostructure.

\section{Conclusions}

The aim of this work has been to investigate theoretically and numerically the electronic structure and 
the superconducting properties of  a nano-structured superlattice of quantum layers in the presence of RSOC. 
We have described the unconventional Lifshitz transition in a 3D superlattice of metallic layers characterized 
by the length of the circular nodal line increasing with RSOC in the negative helicity states of the spin-orbit split electron spectrum. 
Here we have provided the description of the tuning the multigap Bogoliubov superconductivity near the bottom of the upper subband with negative helicity shifted by the RSOC. Our theory overcomes the limitations present so far due to common BCS approximations  used in previous theoretical works on superconductivity in presence of spin-orbit interactions which  mostly describe superconductivity only at  very high Fermi energy.
The work in Re.[\onlinecite{Savary2017}] constitutes an important exception, focussing on superconductivity in low-density semimetals in the presence of strong spin-orbit coupling and analyzing the superconducting instability in different pairing channels. This latter work clearly shows the need to systematically develop the extension of the BCS theory in strongly spin-orbit coupled systems (see also [\onlinecite{brydon2016}]). 
We have shown the key role of the quantum configuration interaction between the gaps
in the self-consistent mean-field equation which requires the calculation of the exchange interactions between singlet pairs in subbands with different quantum number and different helicity. The exchange interactions are key contact interactions which have been shown to be essential in condensation phenomena in fermionic quantum ultracold gases. In our theory the contact interactions are in action together with the phonon exchange cooper pairing. The key result of this work has been  the calculation of the overlap of the electron wave.functions by solving the non relativistic Dirac equation. 
The results of this work provide a roadmap for the quantum material design of a superlattice of periodicity $d$ made of superconducting atomic flakes of thickness $L$ separated by spacers of thickness $W$ where the energy dispersion in the transversal direction is of the order the pairing energy cut-off and the spin-orbit length is of the order of the 3D superlattice period.
Resonant and crossover phenomena in the normal state are amplified when the transverse energy dispersion of electrons in the superlattice is of the same order of magnitude of the energy cutoff $\Delta E_z \sim \hslash \omega_0$ of the effective pairing interaction. Under these conditions the introduction of a RSOC creates a completely unexpected variation in the topology of the Fermi surface, especially for the negative helicity band. In particular, the RSOC induces an unconventional Lifshitz transition with an associated extended van Hove singularity. 
For the non-BCS superconducting phase we have solved the Bogoliubov equation for the multiple gaps numerically. The unusual complexity in the properties of the normal phase is reflected in an amplification of the gap and the critical temperature in precise energy ranges.  
We have found that the enhancement of the superconducting parameters takes place when the chemical potential is tuned around the Lifshitz transition. Under these circumstances it is necessary to include the configuration interaction between different gaps in different subbands.

The issue of superconducting fluctuations in a multiband and multigap configuration deserves a comment at this point.  Whereas amplitude and phase fluctuations of the order parameter are in general detrimental and a source of large suppression of the (otherwise enhanced) critical temperature in low dimensional and/or strongly coupled superconductors,  their effect can be reduced by the recently proposed mechanism [\onlinecite{salasnich2019screening,Saraiva2020}]  of the screening of superconducting fluctuations in a (at least) two-band system. 
References [\onlinecite{salasnich2019screening,Saraiva2020}] demonstrated that a coexistence of a shallow carrier band with strong pairing and a deep band with weak pairing, together with the exchange-like pair transfer between the bands to couple the two condensates, realizes an optimal and robust multicomponent superconductivity regime: 
it preserves strong pairing to generate large gaps and a very high critical temperature but screens the detrimental superconducting fluctuations, thereby suppressing the pseudogap state. The screening is found to be very efficient even when the pair exchange is very small. Thus, a multi-band superconductor with a coherent mixture of condensates in the BCS regime (deep band) and in the BCS-BEC crossover regime (shallow band) offers a promising route to enhance critical temperatures, eliminating at the same time the suppression effect due to fluctuations.
In the light of these considerations, a quantitative calculation of the screening in the system here considered, requiring the inclusion of the spin-orbit coupling terms in the fluctuation propagator,  is postponed to a future work. 

The coexistence of at least one large Fermi surface and at least one small Fermi surface appearing or disappearing with small changes in the chemical potential is the key ingredient for the shape resonance idea in superconducting gaps [\onlinecite{bianconi2005feshbach}, \onlinecite{  innocenti2010resonant}] which is a type of Fano-Feshbach resonance.  By changing the chemical potential, the critical temperature ($T_C$) decreases towards $0$ K when the chemical potential is tuned to the band edge, because of the Fano antiresonance, and the $T_C$ maximum appears (as in Fano resonances) at higher energy, between one and two times the pairing interaction above the band edge [\onlinecite{bianconi2005feshbach}, \onlinecite{innocenti2010resonant}, \onlinecite{bianconi2014shape}]. 
Finally, one of the most interesting aspects highlighted by this work is the existence of an optimal condition for the amplification of the critical temperature when the band shift due to RSOC  is larger than the dispersion along $ z $ of the upper subband and the cut-off energy.

\begin{acknowledgments}
We gratefully thank Andrea Perali for discussions.
We thanks the staff of  Department of Mathematics and Physics of Roma Tre University and Superstripes-onlus for support of this research project.
\end{acknowledgments}	

%\bibliography{Bibliografia} 
\input{Paper_Revised_V2.bbl}
\end{document}